\documentclass{article}
\usepackage{smc}
\usepackage{times}
\usepackage{ifpdf}
\usepackage[english]{babel}
\usepackage{cite}

\def\papertitle{LEARNING SPARSE ANALYTIC FILTERS FOR PIANO TRANSCRIPTION}
\def\firstauthor{Frank Cwitkowitz}
\def\secondauthor{Mojtaba Heydari}
\def\thirdauthor{Zhiyao Duan}

\newif\ifpdf
\ifx\pdfoutput\relax
\else
   \ifcase\pdfoutput
      \pdffalse
   \else
      \pdftrue
\fi

\ifpdf 
  \usepackage[pdftex,
    pdftitle={\papertitle},
    pdfauthor={\firstauthor, \secondauthor, \thirdauthor},
    bookmarksnumbered, 
    pdfstartview=XYZ 
   ]{hyperref}

  \usepackage[pdftex]{graphicx}
  \graphicspath{{./figures/}}
  \DeclareGraphicsExtensions{.pdf,.jpeg,.png}

  \usepackage[figure,table]{hypcap}

\else 
  \usepackage[dvips,
    bookmarksnumbered, 
    pdfstartview=XYZ 
  ]{hyperref}  

  \usepackage[dvips]{epsfig,graphicx}
  \graphicspath{{./figures/}}
  \DeclareGraphicsExtensions{.eps}

  \usepackage[figure,table]{hypcap}
\fi

\hypersetup{
    colorlinks,%
    citecolor=black,%
    filecolor=black,%
    linkcolor=black,%
    urlcolor=black
}

\usepackage{xfp} 
\usepackage{tikz} 
\usetikzlibrary{positioning}

\title{\papertitle}

\oneauthor
   {\firstauthor, \hspace{0.5em} \secondauthor, \hspace{0.5em} \thirdauthor} {Audio Information Research Lab, University of Rochester \\ %
     {\small\tt \(\{\)\href{mailto:fcwitkow@ur.rochester.edu}{fcwitkow},\href{mailto:mheydari@ur.rochester.edu}{mheydari}\(\}\)@ur.rochester.edu \href{mailto:zhiyao.duan@rochester.edu}{zhiyao.duan@rochester.edu}}}

\begin{document}
\capstartfalse
\maketitle
\capstarttrue
\begin{abstract}
In recent years, filterbank learning has become an increasingly popular strategy for various audio-related machine learning tasks. This is partly due to its ability to discover task-specific audio characteristics which can be leveraged in downstream processing. It is also a natural extension of the nearly ubiquitous deep learning methods employed to tackle a diverse array of audio applications. In this work, several variations of a frontend filterbank learning module are investigated for piano transcription, a challenging low-level music information retrieval task. We build upon a standard piano transcription model, modifying only the feature extraction stage. The filterbank module is designed such that its complex filters are unconstrained 1D convolutional kernels with long receptive fields. Additional variations employ the Hilbert transform to render the filters intrinsically analytic and apply variational dropout to promote filterbank sparsity. Transcription results are compared across all experiments, and we offer visualization and analysis of the filterbanks.
\end{abstract}

\section{Introduction}\label{sec:introduction}
Automatic music transcription (AMT) is an essential capability for intelligent systems which analyze music \cite{benetos2018automatic}. The task is part of the broader music information retrieval (MIR) class of machine learning problems. AMT seeks to retrieve all of the information necessary to develop a score which accurately represents the music. However, given this complexity, the problem is typically reduced to the aim of estimating all notes, where a note is characterized by its pitch, time of onset, and duration \cite{benetos2018automatic}. Multi-instrument AMT is challenging, leading many to develop algorithms targeted for successful single-instrument AMT as an initial goal. In particular, piano transcription has been an active research task in AMT, given the wide availability of note annotations \cite{hawthorne2018enabling}, and the consistency of piano timbre compared to multi-instrument ensembles \cite{cogliati2015piano}.

Most machine learning methods involving AMT, and MIR in general, begin with a feature extraction stage which calculates a 2D time-frequency representation (TFR) for a piece of audio. Common choices for this step include the Mel-spectrogram or the constant-Q transform (CQT). In AMT algorithms, the features must contain at least enough information to detect and track notes. While standard TFRs are viable for this purpose, they may not be optimal. In particular, most frequency analysis methods do not explicitly model note characteristics. They also require careful design choices like filter shapes and other hyperparameters. Furthermore, there is fundamentally no information gained in moving to the frequency-domain.

In this work, we focus on filterbank learning for piano transcription, extending the preliminary work of \cite{cwitkowitz2019end}. Filterbank learning is a way to circumvent the use of handcrafted features by replacing or augmenting the TFR calculation with the response from a bank of learnable filters. As the learnable filters are tuned jointly with a backend model for the task at hand, ideally they can model task- and domain-specific characteristics of the input signal.

Although filterbank learning is widely applicable to various audio processing problems, we target piano transcription because it is a complex task which, intuitively, may benefit from modeling notes in the time-domain. The filterbank learning paradigm provides an opportunity to model some of the more obscure properties of musical notes, such as onset or offset behavior, harmonic structure, inharmonicity, or more generally timbre. Piano transcription is also a task with plenty of annotated data and consistency, two characteristics which can simplify the problem.

We adopt the Onsets \& Frames piano transcription model \cite{hawthorne2017onsets, hawthorne2018enabling} and replace the feature extraction stage with a learnable filterbank module\footnote{All code is available at \url{https://github.com/cwitkowitz/sparse-analytic-filters}.}. We utilize a 1D convolutional structure for the filterbank, and experiment with several initialization strategies, model variations and regularization techniques. The filters represented in the module are complex, and can be employed with a standard hop size or stride. We compare the transcription results of each experiment, and provide an analysis of the learned filterbanks. We show that in general, the filters converge to sparse, unique shapes, which we speculate to be modeling various note characteristics. We also demonstrate that filterbanks initialized with random weights achieve comparable performance to those with fixed TFR initializations.

\section{Related Work}\label{sec:related_work}
\begin{figure*}[t]
\centerline{\includegraphics[width=\linewidth]{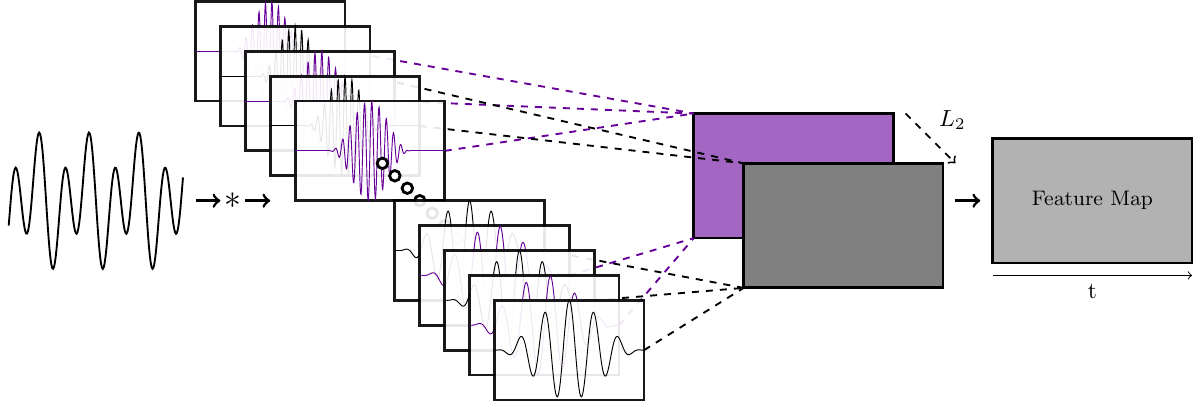}}
\caption{Complex filterbank learning module. The real (black) and imaginary (purple) part of each filter are learned independently. The real and imaginary responses are combined channel-wise using $L_2$ pooling to obtain a feature map.}
\label{fig:classic_diagram}
\end{figure*}

Filterbank learning has gained traction in recent years as a means to perform audio-related machine learning tasks in an end-to-end fashion. Some of the first filterbank learning attempts emerged to tackle problems within the speech community \cite{sainath2015learning, hoshen2015speech}, replacing the more widely used Mel-Frequency Cepstral Coefficients. This eventually led to powerful models for tasks such as speech separation \cite{luo2018tasnet, luo2019conv}.

Filterbank learning has also been applied as a frontend for music data. However, this has mainly been for high-level tasks such as music auto-tagging \cite{dieleman2014end, pons2017end, kim2018sample} or various classification problems \cite{lee2018samplecnn}. For lower-level tasks, such as AMT, filterbank learning has received only some attention \cite{thickstun2016learning, carvalho2017towards}. Music data has rich characteristics, patterns, and relationships at many layers of abstraction, including the note-level, the instrument-level, etc. As such, filterbank learning carries significant potential for discovering, capturing, and leveraging task-specific information.

Many filterbank learning approaches learn strictly real-valued filters, and attain shift-invariance by pooling the response at small hops across time \cite{sainath2015learning}. These filterbanks are analogous to standard 1D convolutional blocks in deep networks. In contrast, the filters used in fixed TFR calculations, e.g., CQT, are typically complex and analytic, which means that they implicitly encode phase. As such, they are stable with relatively large hop sizes, and there is no need to apply further temporal pooling on their responses.

Some filterbank learning approaches have extended the 1D convolutional approach to implement complex filters with grouped real-valued filters representing the real and imaginary part of the complex filter \cite{zeghidour2018learning, zeghidour2018end, lattner2019learning}. While the filters in these examples are complex, if their analytic property is not preserved, they will have an asymmetric frequency response about 0 Hz and no longer exhibit shift-invariance in the magnitude response. Moreover, the 1D convolutional layer approach sometimes suffers from too little constraints, leading to noisy filters with no distinct or localized frequency response. One way to address this is to add soft constraints, e.g., initializing the filterbank with the weights approximating a standard transform \cite{sainath2015learning}.

Recently, there has been a trend to further constrain filterbank learning \cite{ravanelli2018speaker, zeghidour2021leaf}, by learning a small number of parameters which define fixed-shape filters. While these approaches may exhibit more stability, they can only really model simple bandpass filters and thus, lower the potential for discovering meaningful patterns in data at the signal level. Other approaches learn frequency-domain filters on top of spectrograms \cite{won2020data, zhang2019deep}, initializing them with harmonic relationships. We wish to discover these relationships naturally, rather than imposing them as a constraint.

Many approaches to piano transcription have attempted to realize a framework which can learn task-specific note characteristics. Non-Negative Matrix Factorization (NMF) has been proposed to learn properties such as the harmonic relationships, temporal evolution, attack, and decay of piano notes from a TFR \cite{vincent2009adaptive, cheng2016attack}. Convolutional sparse coding is a similar approach which operates in the time-domain, and was proposed as a way to estimate the activation of pre-acquired note impulse responses \cite{cogliati2015piano, cogliati2017piano}.

More recently, deep neural networks (DNNs) have demonstrated success in learning to estimate discrete pitch activity from TFRs \cite{sigtia2016end, hawthorne2017onsets, hawthorne2018enabling, kim2019adversarial}. Some approaches have attempted to design DNNs such that they naturally leverage information about note characteristics \cite{kelz2019deep, kwon2020polyphonic, kong2021high, yan2021skipping}. Since they are very powerful and efficient at learning features for many tasks, we hypothesize that DNNs can be utilized to learn better input features for acoustic models. Furthermore, we believe the proposed frontend will naturally learn to model note characteristics for piano transcription.

\section{Method}
The complex filterbank learning module is implemented as a 1D convolutional layer. It accepts a 1D signal as input and produces a real and imaginary feature map, which are combined using $L_2$ pooling, a simple mechanism for computing the magnitude. The output of the module is subsequently converted to log-amplitude and fed into a batch normalization layer. The filterbank is formulated such that an inner product is taken between a time-domain signal $x$ and $n_{bins}$ filters, indexed by $\mu$, of respective lengths $l_\mu$ with weights $\theta_\mu$ at discrete hops $k$ spaced $l_h$ samples apart:
\begin{equation}
    X[k, \mu] = \sum_{n=0}^{l_{\mu}-1} x[k l_{h} + n] \theta_{\mu}[n].
\end{equation}
Note that this operation is equivalent to convolution, or more precisely correlation, using a stride of $l_h$. The most straightforward way to represent complex filters in this type of framework is to allocate two adjacent filters for the real and imaginary weights, and to combine their respective responses channel-wise using $L_2$ pooling \cite{zeghidour2018learning}:
\begin{equation}
    \left| X[\mu] \right| = \sqrt{\left(x * \mathcal{R}(\theta_\mu)\right)^2 + \left(x * \mathcal{I}(\theta_\mu)\right)^2}.
\end{equation}
Note that with this representation, there are actually $2 \times n_{bins}$ filters to learn, and they are only implicitly associated via the subsequent pooling mechanism. Fig. \ref{fig:classic_diagram} illustrates this filterbank architecture. While this approach can model complex filters, the filters are rarely analytic, unless they are initialized as analytic filters. This means that they may have non-zero responses to negative frequencies and may not be shift-invariant with respect to time.

Another way to model complex filters is to learn only the real part and to infer the imaginary part such that the complex filter is analytic \cite{pariente2020filterbank}. This can be done using the Hilbert Transform, which computes the imaginary counterpart to a real signal, such that the resulting complex signal is analytic\footnote{Note that the filters represented with this variation may only be approximately analytic, due to the limitations of discrete processing.}. This can be expressed as
\begin{equation}
    \left| X[\mu] \right| = \sqrt{\left(x * \mathcal{R}(\theta_\mu)\right)^2 + \left(x * H(\mathcal{R}(\theta_\mu))\right)^2},
\end{equation}
where $H(\cdot)$ denotes the Hilbert transform. This variation learns shift-invariant filters with frequency responses containing energy only in the positive frequency range.

\subsection{Initialization Strategy}
Within the framework presented above, the weights $\theta_\mu$ are initialized randomly by default. Without inserting any prior knowledge into the filterbank, it must learn to generate feature maps from scratch. This strategy has the potential to discover new filter shapes and characteristics that are not present in standard TFRs.

Alternatively, it is possible to insert weights into the filterbank such that it implements a time-domain variable-Q transform (VQT) if left untrained. The complex variable-Q response for filter $\mu$ can be computed in the time-domain by making the weights $\theta_\mu$ complex basis functions with center frequency $f_{\mu}$, sampling rate $f_s$, and smoothly varied Q-factor $Q_{\mu}$:
\begin{equation}
    \theta_{\mu,n} = w_{\mu,n} e^{-j \frac{2 \pi f_{\mu} n}{f_s}}, \hspace{1em} n = 0, \dots, l_{\mu}.
\end{equation}
The filter length $l_{\mu} = \left \lceil Q_{\mu} \frac{f_s}{f_{\mu}} \right \rceil$ is set such that the desired Q-factor, whether constant or smoothly varied, is maintained across all filters. The windowing function $w$ is chosen as a Hanning window and matches the filter length $l_{\mu}$ for each filter. We avoid normalizing the response of each basis by filter length $l_{\mu}$ so that the weights of each filter are all comparable and responsive to a single learning rate. The center frequency of each filter in the VQT initialization is defined by selecting $f_{min}$, and applying
\begin{equation}
    f_{\mu} = f_{min} \times 2^{\frac{\mu}{n_{bpo}}},
\end{equation}
where $n_{bpo}$ is the number of bins or per octave.

The bandwidth of each respective filter is determined by $B_{\mu} = f_{\mu+1} - f_{\mu} + \gamma$. In a CQT, the ratio between the center frequency $f_{\mu}$ and the bandwidth $B_{\mu}$, or the Q-factor, is constant, meaning that $\gamma = 0$. However, for large $n_{bpo}$, this can lead to filters with extremely tight bands, requiring long impulse responses. Modestly utilizing the constant bandwidth offset $\gamma$ relaxes the constancy constraint in the lower frequency range and provides direct control over the size of the filter receptive fields, which must all be zero-padded to the largest filter in order to be stored in a 1D convolutional layer. We also use the VQT parameters to determine the receptive field size for random initialization.

Initializing the filterbank with the VQT weights provides a foundation to improve upon, and could even be thought of as pre-training the filterbank. Although there are multiple transforms one could utilize, the VQT is chosen because it is intuitively a better starting point for modeling note characteristics. Every $\frac{n_{bpo}}{12}^{th}$ filter is already locked onto a pitch in the Western music scale. Since we are primarily concerned with relationships between pitch and temporal characteristics for the AMT problem, we believe this is an appropriate initialization strategy.

An additional initialization strategy is the harmonic comb, where a set of harmonics $\mathcal{H}$ is defined, and separate filters with center frequency $f = h \times f_{\mu}$ are constructed for each $h \in \mathcal{H}$. All filters associated with a given $\mu$ are summed and collapsed into a single set of normalized weights for insertion into the 1D convolutional layer. In this way, each filter responds directly to the harmonics of its corresponding pitch, and characteristics such as inharmonicity or relative harmonic strength (timbre) can be fine-tuned.

\begin{table*}[ht]
\begin{center}
\small
  \begin{tabular}{|c||c|c|c||c|c|c|}
    \hline
    {} & \multicolumn{3}{c||}{\textbf{MAESTRO}} & \multicolumn{3}{c|}{\textbf{MAPS}} \\
    \hline
    {\textbf{Experiment}} & {\textbf{Frame} $\mathbf{F_1}$} & {\textbf{Note-On} $\mathbf{F_1}$} & {\textbf{Note-Off} $\mathbf{F_1}$} & {\textbf{Frame} $\mathbf{F_1}$} & {\textbf{Note-On} $\mathbf{F_1}$} &  {\textbf{Note-Off} $\mathbf{F_1}$} \\
    \hline
    \hline
    \textit{mel} & 91.80* & 95.95* & 83.44* & 81.40* & 81.42* & 59.15* \\
    \hline
    \textit{mel} & \textbf{90.91} & \textbf{95.82} & \textbf{83.14} & \textbf{81.26} & 83.86 & \textbf{59.07} \\
    \hline
    \textit{cqt} & 90.79 & 95.29 & 82.30 & 77.46 & 82.35 & 52.18 \\
    \hline
    \textit{vqt} & 90.18 & 94.74 & 80.51 & 80.26 & 83.42 & 55.34 \\
    \hline
    \textit{fixed comb} & 87.59 & 92.19 & 75.09 & 80.30 & \textbf{84.64} & 57.22 \\
    \hline
    \hline
    \textit{cl+rnd} & 86.70 & 91.22 & 73.72 & 70.03 & 78.77 & 43.92 \\
    \hline
    \textit{cl+vqt} & 87.53 & 92.24 & 75.64 & 72.90 & 80.90 & 48.06 \\
    \hline
    \textit{hb+rnd} & 86.50 & 91.30 & 73.19 & 76.13 & 80.00 & 51.88 \\
    \hline
    \textit{hb+vqt} & \textbf{87.81} & \textbf{92.63} & \textbf{76.01} & 75.11 & 80.71 & 50.66 \\
    \hline
    \textit{hb+rnd+brn} & 85.23 & 90.19 & 70.75 & 74.81 & 79.48 & 50.57 \\
    \hline
    \textit{hb+rnd+gau} & 85.63 & 90.41 & 71.81 & 75.58 & 80.44 & 51.28 \\
    \hline
    \textit{hb+rnd+var} & 86.04 & 90.61 & 72.27 & 73.59 & 80.09 & 47.99 \\
    \hline
    \textit{hb+vqt+var} & 87.45 & 92.39 & 74.92 & 75.62 & 80.80 & 50.55 \\
    \hline
    \textit{hb+comb+var} & 87.52 & 92.27 & 75.69 & \textbf{76.74} & \textbf{80.98} & \textbf{52.57} \\
    \hline
  \end{tabular}
\end{center}
\caption{Evaluation results for all baseline and filterbank experiments (separated by break). Filterbank learning experiment names are formatted in terms of variant (classic (\textit{cl}) / Hilbert (\textit{hb})) \textit{+} initialization (random (\textit{rnd}) / \textit{vqt} / \textit{comb}) \textit{+} dropout (none / Bernoulli (\textit{brn}) / Gaussian (\textit{gau}) / variational (\textit{var})). Results obtained using MAESTRO $V_1$ are indicated by *.}
\label{tab:results}
\end{table*}

\subsection{Variational Dropout}
Variational dropout \cite{molchanov2017variational} is a regularization technique which allows for the learning of sparse frontend filters with long receptive fields. It is similar to Gaussian dropout, except the dropout rate, or variance, corresponding to each weight is learned. During training we treat the weights as random variables, with their true values $\theta$ being the mean, and an additional set of learned parameters $\sigma^2$ being their variance. By learning the variance of each weight, variational dropout induces sparsity, as it pushes less important weights to have higher variance, and more important weights to have lower variance. This leads to more sparse and interpretable filters, and can deal with very long receptive fields effectively. The stochasticity can also stimulate the filterbank to take interesting and unexpected shapes as it is being trained. During each forward pass, noise is sampled from the variance of the filterbank response and added to the mean response:
\begin{equation}
    X[\mu] \sim \mathcal{N}(x * \theta_\mu, x^2 * \sigma_\mu^2).
\end{equation}
The parameters $\sigma_{\mu}^2$ are jointly trained with the filterbank parameters $\theta_{\mu}$, and their gradient is calculated using a very close approximation to the KL-divergence \cite{molchanov2017variational}. In practice, we iterate upon $\log \sigma^2$ for improved stability. It should be noted that we only add noise to the real part of the feature map when using the Hilbert filterbank variant.

\section{Experiments}
\begin{figure*}[!h]
\centerline{\includegraphics[width=1.05\linewidth]{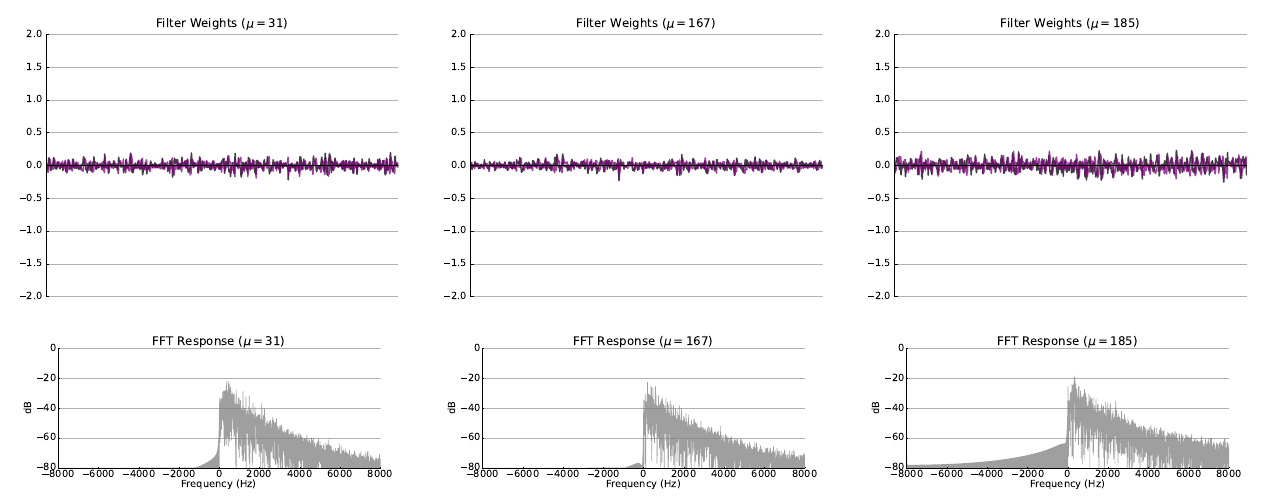}}
\caption{Examples of filters from the \textit{hb+rnd+var} experiment which exhibit a high degree of sparsity. These filters have small weights and as such their responses likely have little relevance to the downstream model.}
\label{fig:hb_rnd_var_sparse_filters}
\end{figure*}

\begin{figure*}[!h]
\centerline{\includegraphics[width=1.05\linewidth]{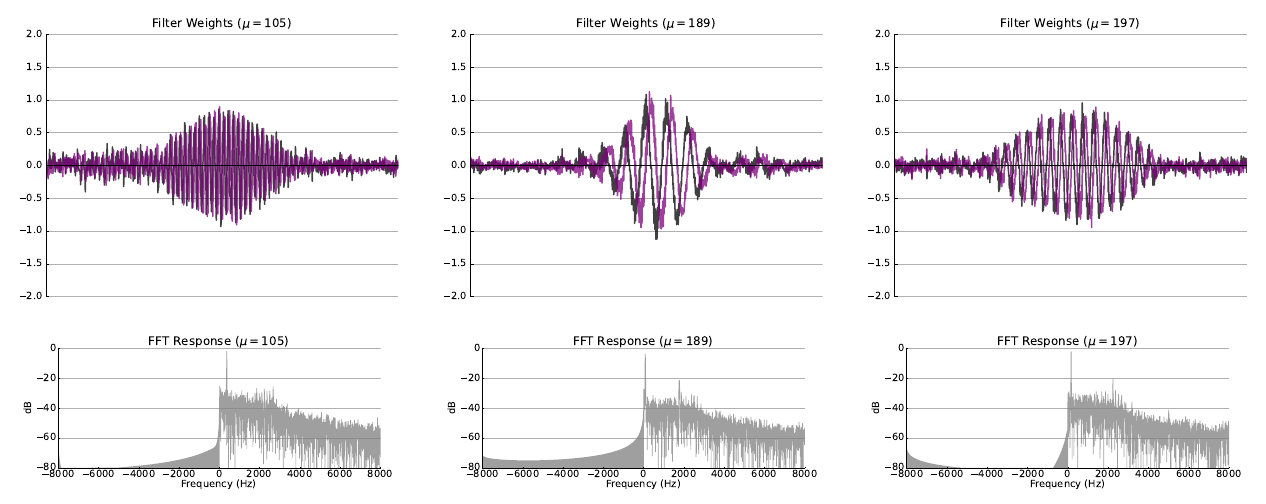}}
\caption{Examples of filters from the \textit{hb+rnd+var} experiment which are well-localized within the receptive field and exhibit coherent shapes in both the time- and frequency-domain.}
\label{fig:hb_rnd_var_localized_filters}
\end{figure*}

\begin{figure*}[!h]
\centerline{\includegraphics[width=1.05\linewidth]{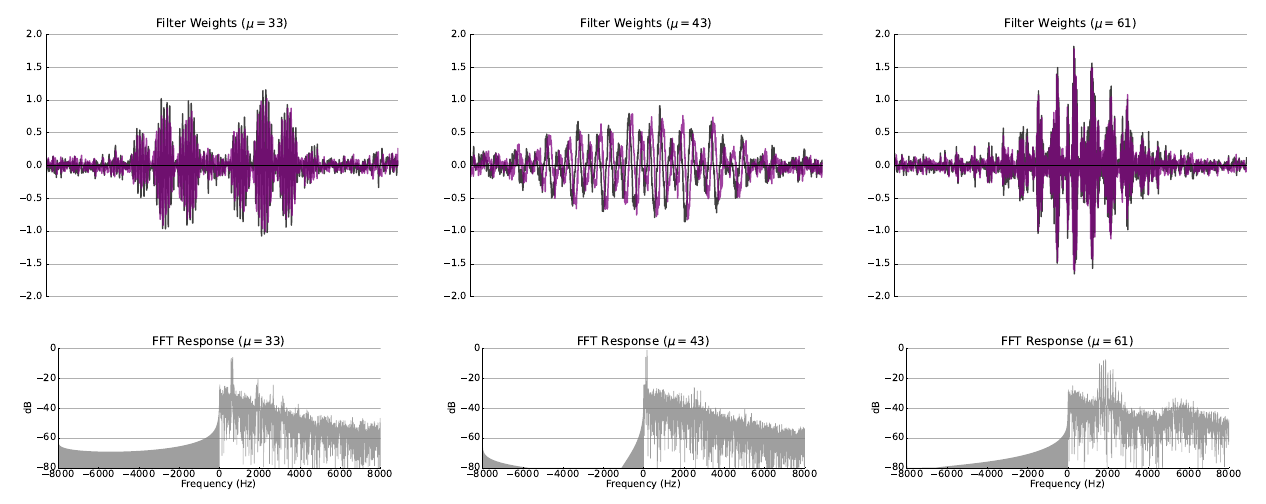}}
\caption{Examples of filters from the \textit{hb+rnd+var} experiment which support multiple frequencies.}
\label{fig:hb_rnd_var_harmonized_filters}
\end{figure*}

\begin{figure*}[!h]
\centerline{\includegraphics[width=1.05\linewidth]{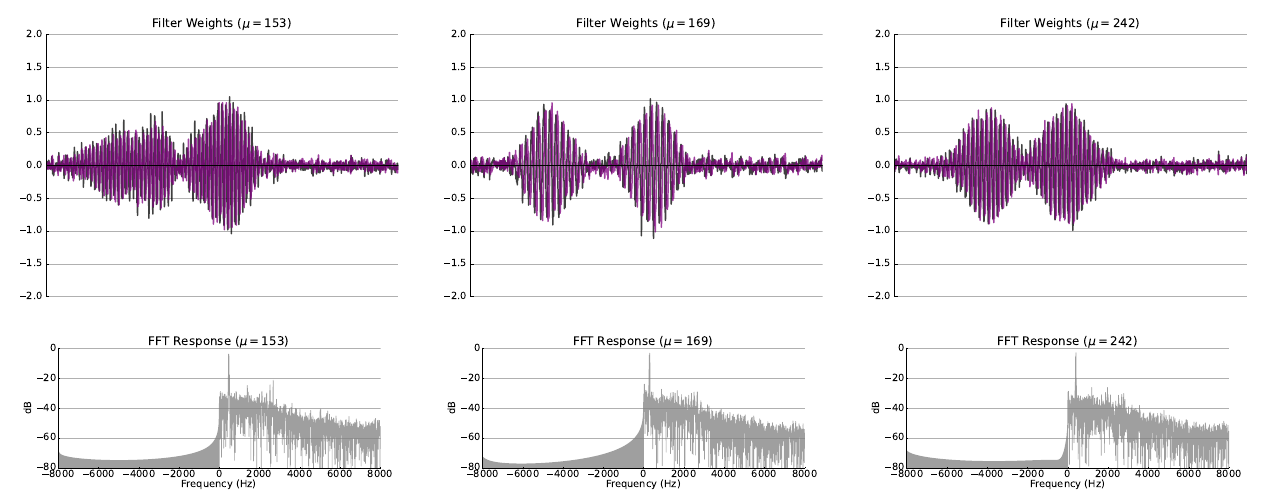}}
\caption{Examples of near-symmetric filters from the \textit{hb+rnd+var} experiment comprising two impulse lobes.}
\label{fig:hb_rnd_var_note_filters}
\end{figure*}

\begin{figure*}[!h]
\centerline{\includegraphics[width=1.05\linewidth]{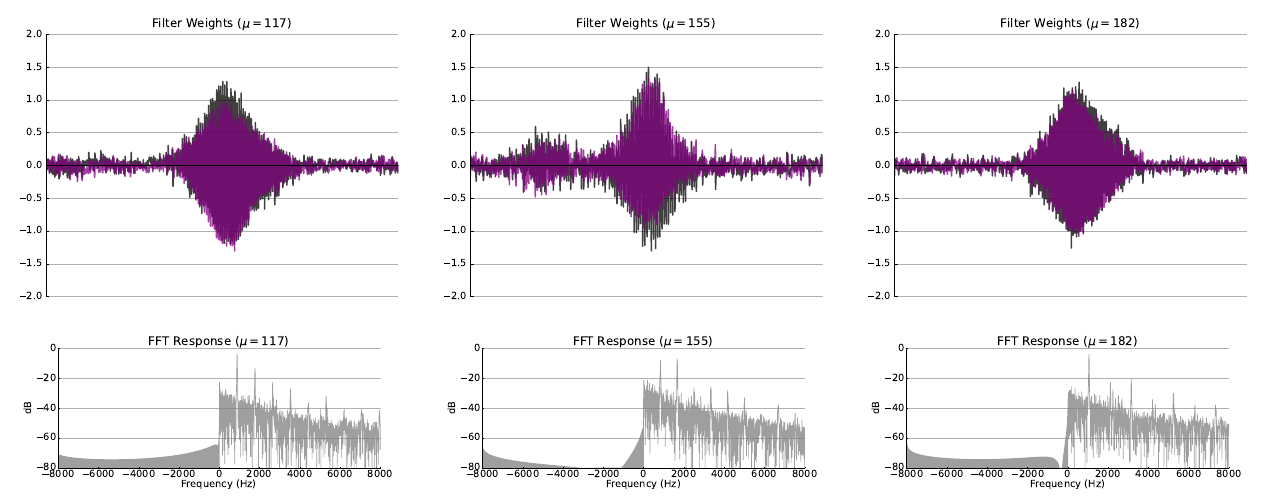}}
\caption{Examples of filters from the \textit{hb+rnd+var} experiment which exhibit  harmonic structure.}
\label{fig:hb_rnd_var_harmonic_filters}
\end{figure*}

\subsection{Model}
The filterbank learning module is used as a frontend replacement for the feature extraction stage of the Onsets \& Frames model \cite{hawthorne2017onsets, hawthorne2018enabling} and jointly trained for the task of piano transcription. While there have been more recent models with improved performance, \textit{e.g.} \cite{yan2021skipping, hawthorne2021sequence}, we stick with Onsets \& Frames as a baseline due to its simplicity for the purposes of filterbank analysis. The original model takes as input a Mel spectrogram and produces a frame-level salience estimate for all note pitches, independently. Note predictions are generated using the simple post-processing steps proposed in the original paper, except we do not use the note predictions to refine the salience estimate, nor do we perform the additional velocity estimation. Our experimental setup is the same as the original paper \cite{hawthorne2017onsets} with the subsequent improvements proposed in 
\cite{hawthorne2018enabling}.
We utilize the same hyperparameters, running each experiment for 2000 iterations, where each music piece is accessed once per iteration to sample a sequence of frames for a batch.

\subsection{Datasets}
MAESTRO \cite{hawthorne2018enabling} is used for training, validation, and testing. Specifically, we use version 3 ($V_3$) of the dataset for all but one experiment, which is run on version 1 ($V_1$). We follow the suggested partitioning for both versions. We also evaluate on the acoustic partitions (ENSTDkAm/ENSTDkCl) of MAPS \cite{emiya2009multipitch}, to inspect the generalization potential of our method. 
We downsample all audio to $f_s = 16$ kHz, and use a hop length of $L_h = 512$ samples.

\subsection{Metrics}
We evaluate experiments using the same transcription metrics as in \cite{hawthorne2017onsets}, which include frame-wise, note-wise, and note-wise with offset scores. Precision, recall, and f1-score are calculated for each of these metrics using \textit{mir\_eval} \cite{raffel2014mir_eval}. For frame-wise evaluation, detected pitches are compared to the ground-truth at the frame-level. For note-wise evaluation, note estimates for an entire piece are compared to the ground-truth. A note is considered correct if its pitch is within half a semitone of the true value, and if its onset is within 50 ms of the true value. The note-wise with offset metric additionally compares the estimated offset to the true value, a correct estimate being within 50 ms or 20\% of the ground truth duration, whichever is larger.

\subsection{Results}
We conduct several experiments to verify our implementation of Onsets \& Frames, and to assess the strength of various features. We use $n_{bins} = 229$ and HTK frequency spacing for Mel spectrogram, as in \cite{hawthorne2018enabling}. We choose $f_{min} \approx 32.7$ Hz, $n_{bins} = 252$ and $n_{bpo} = 36$, or 3 bins per semitone for CQT and VQT. We calculate $\gamma = \frac{24.7}{0.108 * Q}$, such that all filters have a bandwidth proportional to the ERB scale by a constant factor \cite{schorkhuber2014matlab} for the VQT experiment. We also experiment with an untrained filterbank initialized using the harmonic comb strategy ($\mathcal{H} = \{1, ..., 5\}$). Results for these experiments are provided at the top of \tabref{tab:results}. We observe that Mel spectrogram performs best in every metric besides note-onsets for MAPS, where the untrained comb filterbank slightly outperforms it. Interestingly, the comb filterbank is inferior in all other metrics.

Next we conduct several filterbank learning experiments where the 
output of the filterbank is fed to the model instead of a TFR. The same VQT parameters used in the baseline experiments are used to initialize the filterbanks. We vary the filterbank variant, initialization, and regularization method across all experiments. For Gaussian and variational dropout, we initialize all $\log \sigma_{\mu,n}^2 = -10$, and for Bernoulli dropout we use a dropout rate of 10\%. For variational dropout, the KL-divergence term is scaled by a factor of 0.01, such that the implicit sparsity objective does not dominate the training process. Results for these filterbank experiments are provided at the bottom of \tabref{tab:results}.

The Hilbert transform-based architecture performs similarly to the classic framework for random initialization, but performs slightly better for VQT initialization, which tends to slightly outperform the random initialization. The variational dropout regularization method performs the best among regularization methods on MAESTRO, though still worse than experiments with no regularization. None of the learned filterbanks outperform the fixed TFRs. Nonetheless, they demonstrate reasonable performance, even when initialized randomly. The learned filterbanks, however, do not seem to generalize very well to the MAPS dataset. In terms of generalization to MAPS, the harmonic comb initialization does the best, and variational dropout generally underperforms other regularization methods.

\section{Discussion}
We provide visualization of a few examples from the filterbank learning experiment corresponding to the Hilbert variant with random initialization and variational dropout (\textit{hb+rnd+var}), which we consider to have produced the most interesting filters. These are presented in Fig. 2-6. In each figure, the time-domain visualization for three filters is displayed in the top row, with the corresponding frequency-domain visualization for each respective filter in the bottom row. The specific examples are chosen because they appear to belong to the same category of learned filter.

In general, the filters learned for the \textit{hb+rnd+var} experiment tend to exhibit a high degree of sparsity, with coherent shapes in both the time- and frequency-domain. Fig. \ref{fig:hb_rnd_var_sparse_filters} and Fig. \ref{fig:hb_rnd_var_localized_filters} illustrate these properties, respectively. The non-zero filter weights tend to be well-localized within the receptive field, with weights closer to zero being closer to the edges and more prominent weights being closer to the center. The tendency of the filters to be sparse is highly desirable, as it encourages more modular filters and leads to more interpretable shapes in time and frequency. Indeed, it appears that the relevance of a filter can be directly estimated from its amplitude, with less important or even unused filters consisting of mostly near-zero weights rather than random noise. Furthermore, all of the Hilbert variant filters remain approximately analytic throughout the training process. We do not view the artifacts in the negative frequency range to be problematic, and believe they had a negligible effect on the outcome of experiments.

Some additional interesting qualities of the filters are observed, which are presumably related to the modeling of domain-specific characteristics for piano transcription. The spectrum of the filters in Fig. \ref{fig:hb_rnd_var_harmonized_filters} include multiple prominent fundamental frequencies in clusters, which suggests the filters may be modeling harmony or polyphonic piano sounds. The filters in Fig. \ref{fig:hb_rnd_var_note_filters} are nearly symmetric, with two lobes in the impulse response that have overlapping or very close frequency peaks. These filters may be modeling the attack and decay of a piano note with the same pitch. Some of these filters, in addition to the filters in Fig. \ref{fig:hb_rnd_var_harmonic_filters}, exhibit harmonic patterns in the frequency spectrum. In particular, odd harmonics tend to be emphasized in these filters. These observations suggest that the proposed filterbank learning approach is capable of capturing lower-level features which are relevant for music transcription, and that the approach is likely extensible to similar tasks where this type of information is valuable.


The reduced performance of the filterbank learning models can potentially be attributed to too little training or inadequate tuning of hyperparameters. We also speculate that the consistency of piano data, coupled with large model complexity, may have led to overfitting, as evidenced by the lack of generalization to the MAPS dataset. State-of-the-art methods with no filterbank learning already perform piano transcription with reasonable proficiency. As such, it may be that a similar filterbank learning approach would show much more promise in areas with less data and less consistency, such as drum or guitar transcription.

Finally, we stress that the filterbank learning method described in this paper is not presented as a competitive alternative to state-of-the-art piano transcription methodologies. 
We do not offer a comprehensive comparison to such approaches for this reason, but do acknowledge that there are many better methods for the purpose of piano transcription. One may question the utility or validity of the filterbank learning methods due to their reduced performance. While superior performance would have been a nice outcome, we are still quite impressed with the performance of the filterbank learning models, especially the ones initialized randomly. From this perspective, it is clear the proposed methods, \textit{i.e.}, filterbank learning, are successful for piano transcription, albeit not at the same level of performance as standard TFRs like CQT or Mel-spectrogram.

\section{Conclusion}
A filterbank learning module and various techniques were analyzed for the task of piano transcription. 
There are several reasons to suggest that the learned filters model note characteristics such as attack and decay, harmonic structure and relative strength, or inharmonicity. Given these are just hypotheses, a more objective explanation and analysis of the characteristics of the filters learned with our methods is left to future work.
We showed it is possible to learn sparse filters from scratch, and that resulting filters differ significantly from those of the fixed transforms. Although the learned filterbanks did not surpass the performance of the fixed transforms, our method shows promise in learning to model time-domain note characteristics, and demonstrates that randomly initialized filterbanks can perform comparably to standard TFR initializations.

\begin{acknowledgments}
This work has been funded by the National Science Foundation grants IIS-1846184 and DGE-1922591.
We would also like to thank Dr. Juan Cockburn and Dr. Andres Kwasinski for their guidance during preliminary work \cite{cwitkowitz2019end}.
\end{acknowledgments} 

\bibliography{paper}

\begin{thebibliography}{10}
\providecommand{\url}[1]{#1}
\csname url@samestyle\endcsname
\providecommand{\newblock}{\relax}
\providecommand{\bibinfo}[2]{#2}
\providecommand{\BIBentrySTDinterwordspacing}{\spaceskip=0pt\relax}
\providecommand{\BIBentryALTinterwordstretchfactor}{4}
\providecommand{\BIBentryALTinterwordspacing}{\spaceskip=\fontdimen2\font plus
\BIBentryALTinterwordstretchfactor\fontdimen3\font minus
  \fontdimen4\font\relax}
\providecommand{\BIBforeignlanguage}[2]{{%
\expandafter\ifx\csname l@#1\endcsname\relax
\typeout{** WARNING: IEEEtran.bst: No hyphenation pattern has been}%
\typeout{** loaded for the language `#1'. Using the pattern for}%
\typeout{** the default language instead.}%
\else
\language=\csname l@#1\endcsname
\fi
#2}}
\providecommand{\BIBdecl}{\relax}
\BIBdecl

\bibitem{benetos2018automatic}
E.~Benetos, S.~Dixon, Z.~Duan, and S.~Ewert, ``Automatic music transcription:
  An overview,'' \emph{IEEE Signal Processing Magazine}, vol.~36, no.~1, pp.
  20--30, 2019.

\bibitem{hawthorne2018enabling}
C.~Hawthorne, A.~Stasyuk, A.~Roberts, I.~Simon, C.-Z.~A. Huang, S.~Dieleman,
  E.~Elsen, J.~Engel, and D.~Eck, ``Enabling factorized piano music modeling
  and generation with the {MAESTRO} dataset,'' in \emph{Proceedings of ICLR},
  2019.

\bibitem{cogliati2015piano}
A.~Cogliati, Z.~Duan, and B.~Wohlberg, ``Piano music transcription with fast
  convolutional sparse coding,'' in \emph{IEEE 25th International Workshop on
  Machine Learning for Signal Processing (MLSP)}, 2015.

\bibitem{cwitkowitz2019end}
F.~Cwitkowitz, ``End-to-end music transcription using fine-tuned variable-{Q}
  filterbanks,'' Master's thesis, Rochester Institute of Technology, 2019.

\bibitem{hawthorne2017onsets}
C.~Hawthorne, E.~Elsen, J.~Song, A.~Roberts, I.~Simon, C.~Raffel, J.~Engel,
  S.~Oore, and D.~Eck, ``Onsets and frames: Dual-objective piano
  transcription,'' in \emph{Proceedings of ISMIR}, 2018.

\bibitem{sainath2015learning}
T.~N. Sainath, R.~J. Weiss, A.~Senior, K.~W. Wilson, and O.~Vinyals, ``Learning
  the speech front-end with raw waveform {CLDNNs},'' in \emph{Proceedings of
  Interspeech}, 2015.

\bibitem{hoshen2015speech}
Y.~Hoshen, R.~J. Weiss, and K.~W. Wilson, ``Speech acoustic modeling from raw
  multichannel waveforms,'' in \emph{Proceedings of ICASSP}, 2015.

\bibitem{luo2018tasnet}
Y.~Luo and N.~Mesgarani, ``{TasNet}: Time-domain audio separation network for
  real-time, single-channel speech separation,'' in \emph{Proceedings of
  ICASSP}, 2018.

\bibitem{luo2019conv}
------, ``{Conv-TasNet}: Surpassing ideal time–frequency magnitude masking
  for speech separation,'' \emph{IEEE/ACM Transactions on Audio, Speech, and
  Language Processing (TASLP)}, vol.~27, no.~8, pp. 1256--1266, 2019.

\bibitem{dieleman2014end}
S.~Dieleman and B.~Schrauwen, ``End-to-end learning for music audio,'' in
  \emph{Proceedings of ICASSP}, 2014.

\bibitem{pons2017end}
J.~Pons, O.~Nieto, M.~Prockup, E.~Schmidt, A.~Ehmann, and X.~Serra,
  ``End-to-end learning for music audio tagging at scale,'' in
  \emph{Proceedings of ISMIR}, 2018.

\bibitem{kim2018sample}
T.~Kim, J.~Lee, and J.~Nam, ``Sample-level {CNN} architectures for music
  auto-tagging using raw waveforms,'' in \emph{Proceedings of ICASSP}, 2018.

\bibitem{lee2018samplecnn}
J.~Lee, J.~Park, K.~L. Kim, and J.~Nam, ``Sample{CNN}: End-to-end deep
  convolutional neural networks using very small filters for music
  classification,'' \emph{Applied Sciences}, vol.~8, no.~1, 2018.

\bibitem{thickstun2016learning}
J.~Thickstun, Z.~Harchaoui, and S.~Kakade, ``Learning features of music from
  scratch,'' in \emph{Proceedings of ICLR}, 2017.

\bibitem{carvalho2017towards}
R.~G.~C. Carvalho and P.~Smaragdis, ``Towards end-to-end polyphonic music
  transcription: Transforming music audio directly to a score,'' in \emph{IEEE
  Workshop on Applications of Signal Processing to Audio and Acoustics
  (WASPAA)}, 2017.

\bibitem{zeghidour2018learning}
N.~Zeghidour, N.~Usunier, I.~Kokkinos, T.~Schaiz, G.~Synnaeve, and E.~Dupoux,
  ``Learning filterbanks from raw speech for phone recognition,'' in
  \emph{Proceedings of ICASSP}, 2018.

\bibitem{zeghidour2018end}
N.~Zeghidour, N.~Usunier, G.~Synnaeve, R.~Collobert, and E.~Dupoux,
  ``End-to-end speech recognition from the raw waveform,'' in \emph{Proceedings
  of Interspeech}, 2018.

\bibitem{lattner2019learning}
S.~Lattner, M.~D{\"o}rfler, and A.~Arzt, ``Learning complex basis functions for
  invariant representations of audio,'' in \emph{Proceedings of ISMIR}, 2019.

\bibitem{ravanelli2018speaker}
M.~Ravanelli and Y.~Bengio, ``Speaker recognition from raw waveform with
  {SincNet},'' in \emph{IEEE Spoken Language Technology Workshop (SLT)}, 2018.

\bibitem{zeghidour2021leaf}
N.~Zeghidour, O.~Teboul, F.~d.~C. Quitry, and M.~Tagliasacchi, ``{LEAF}: A
  learnable frontend for audio classification,'' in \emph{Proceedings of ICLR},
  2021.

\bibitem{won2020data}
M.~Won, S.~Chun, O.~Nieto, and X.~Serrc, ``Data-driven harmonic filters for
  audio representation learning,'' in \emph{Proceedings of ICASSP}, 2020.

\bibitem{zhang2019deep}
Z.~Zhang, Y.~Wang, C.~Gan, J.~Wu, J.~B. Tenenbaum, A.~Torralba, and W.~T.
  Freeman, ``Deep audio priors emerge from harmonic convolutional networks,''
  in \emph{Proceedings of ICLR}, 2020.

\bibitem{vincent2009adaptive}
E.~Vincent, N.~Bertin, and R.~Badeau, ``Adaptive harmonic spectral
  decomposition for multiple pitch estimation,'' \emph{IEEE Transactions on
  Audio, Speech, and Language Processing (TASLP)}, vol.~18, no.~3, pp.
  528--537, 2010.

\bibitem{cheng2016attack}
T.~Cheng, M.~Mauch, E.~Benetos, S.~Dixon \emph{et~al.}, ``An attack/decay model
  for piano transcription,'' in \emph{Proceedings of ISMIR}, 2016.

\bibitem{cogliati2017piano}
A.~Cogliati, Z.~Duan, and B.~Wohlberg, ``Piano transcription with convolutional
  sparse lateral inhibition,'' \emph{IEEE Signal Processing Letters}, vol.~24,
  no.~4, pp. 392--396, 2017.

\bibitem{sigtia2016end}
S.~Sigtia, E.~Benetos, and S.~Dixon, ``An end-to-end neural network for
  polyphonic piano music transcription,'' \emph{IEEE/ACM Transactions on Audio,
  Speech, and Language Processing (TASLP)}, vol.~24, no.~5, pp. 927--939, 2016.

\bibitem{kim2019adversarial}
J.~W. Kim and J.~P. Bello, ``Adversarial learning for improved onsets and
  frames music transcription,'' in \emph{Proceedings of ISMIR}, 2019.

\bibitem{kelz2019deep}
R.~Kelz, S.~Böck, and G.~Widmer, ``Deep polyphonic {ADSR} piano note
  transcription,'' in \emph{Proceedings of ICASSP}, 2019.

\bibitem{kwon2020polyphonic}
T.~Kwon, D.~Jeong, and J.~Nam, ``Polyphonic piano transcription using
  autoregressive multi-state note model,'' in \emph{Proceedings of ISMIR},
  2020.

\bibitem{kong2021high}
Q.~Kong, B.~Li, X.~Song, Y.~Wan, and Y.~Wang, ``High-resolution piano
  transcription with pedals by regressing onset and offset times,''
  \emph{IEEE/ACM Transactions on Audio, Speech, and Language Processing
  (TASLP)}, vol.~29, pp. 3707--3717, 2021.

\bibitem{yan2021skipping}
Y.~Yan, F.~Cwitkowitz, and Z.~Duan, ``Skipping the frame-level: Event-based
  piano transcription with neural semi-{CRF}s,'' in \emph{Advances in Neural
  Information Processing Systems 34 (NeurIPS)}, 2021.

\bibitem{pariente2020filterbank}
M.~Pariente, S.~Cornell, A.~Deleforge, and E.~Vincent, ``Filterbank design for
  end-to-end speech separation,'' in \emph{Proceedings of ICASSP}, 2020.

\bibitem{molchanov2017variational}
D.~Molchanov, A.~Ashukha, and D.~Vetrov, ``Variational dropout sparsifies deep
  neural networks,'' in \emph{Proceedings of ICML}, 2017.

\bibitem{hawthorne2021sequence}
C.~Hawthorne, I.~Simon, R.~Swavely, E.~Manilow, and J.~Engel,
  ``Sequence-to-sequence piano transcription with transformers,'' in
  \emph{Proceedings of ISMIR}, 2021.

\bibitem{emiya2009multipitch}
V.~Emiya, R.~Badeau, and B.~David, ``Multipitch estimation of piano sounds
  using a new probabilistic spectral smoothness principle,'' \emph{IEEE
  Transactions on Audio, Speech, and Language Processing (TASLP)}, vol.~18,
  no.~6, pp. 1643--1654, 2010.

\bibitem{raffel2014mir_eval}
C.~Raffel, B.~McFee, E.~J. Humphrey, J.~Salamon, O.~Nieto, D.~Liang, D.~P.
  Ellis, and C.~C. Raffel, ``mir\_eval: A transparent implementation of common
  {MIR} metrics,'' in \emph{Proceedings of ISMIR}, 2014.

\bibitem{schorkhuber2014matlab}
C.~Sch{\"o}rkhuber, A.~Klapuri, N.~Holighaus, and M.~D{\"o}rfler, ``A matlab
  toolbox for efficient perfect reconstruction time-frequency transforms with
  log-frequency resolution,'' in \emph{Audio Engineering Society (AES) 53rd
  Conference on Semantic Audio}, 2014.

\end{thebibliography}

\pagebreak
\onecolumn
\section{Appendix}
Here we provide some additional visualization for the filters learned during auxiliary experiments. The visualization is by no means exhaustive. However, the exemplified filters should be sufficient as representatives for each respective experiment.

\subsection{Hilbert + Random + Variational (hb+rnd+var)}
\begin{figure}[!h]
\includegraphics[width=\textwidth]{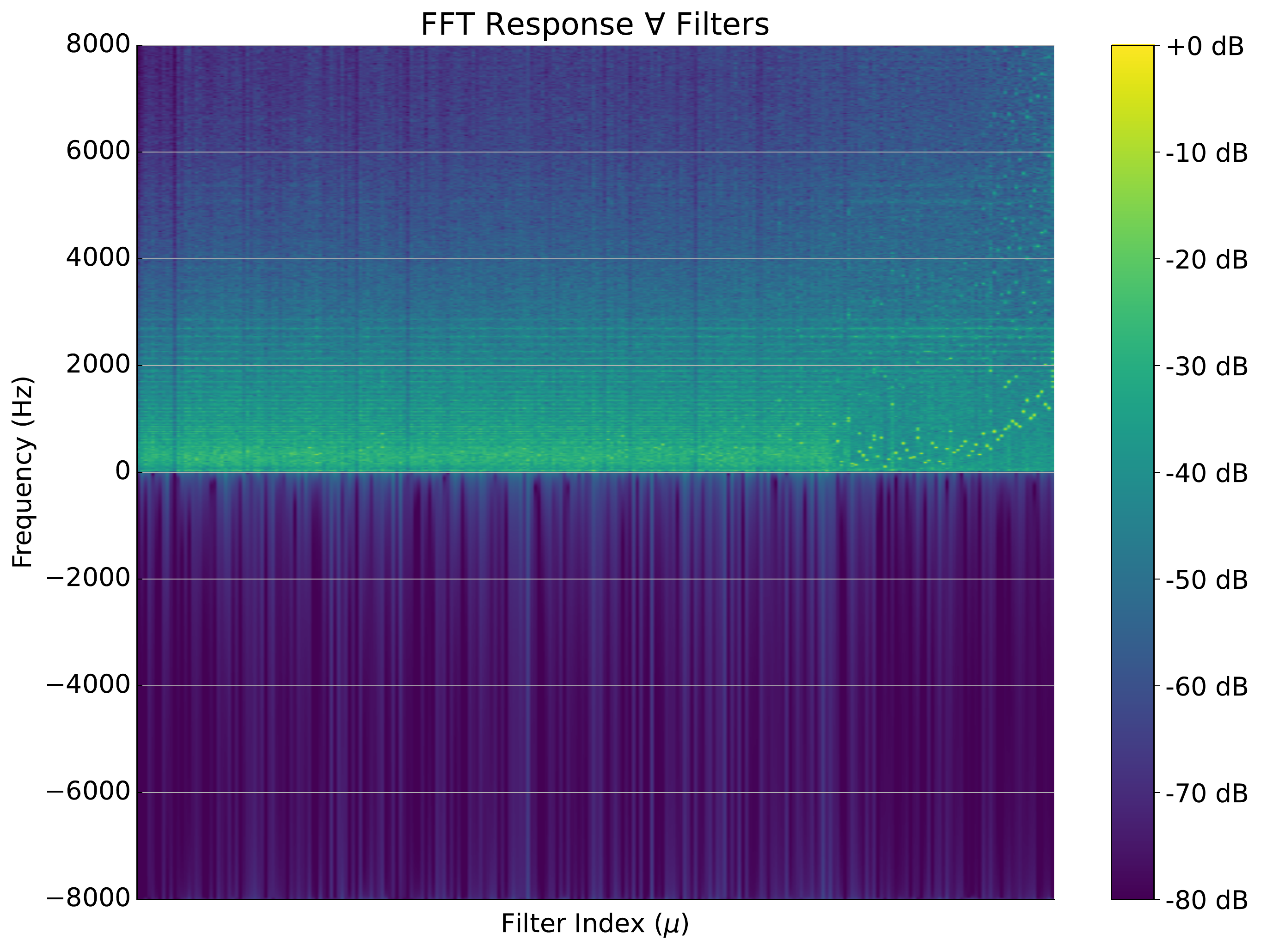}
\caption{Frequency response for all filters of the \textit{hb+rnd+var} (main) experiment, ordered by spectral centroid.}
\label{fig:hb_rnd_var_filters}
\end{figure}

\vfill

\clearpage

\subsection{Classic + Random (cl+rnd)}
\begin{figure}[!h]
\includegraphics[width=\textwidth]{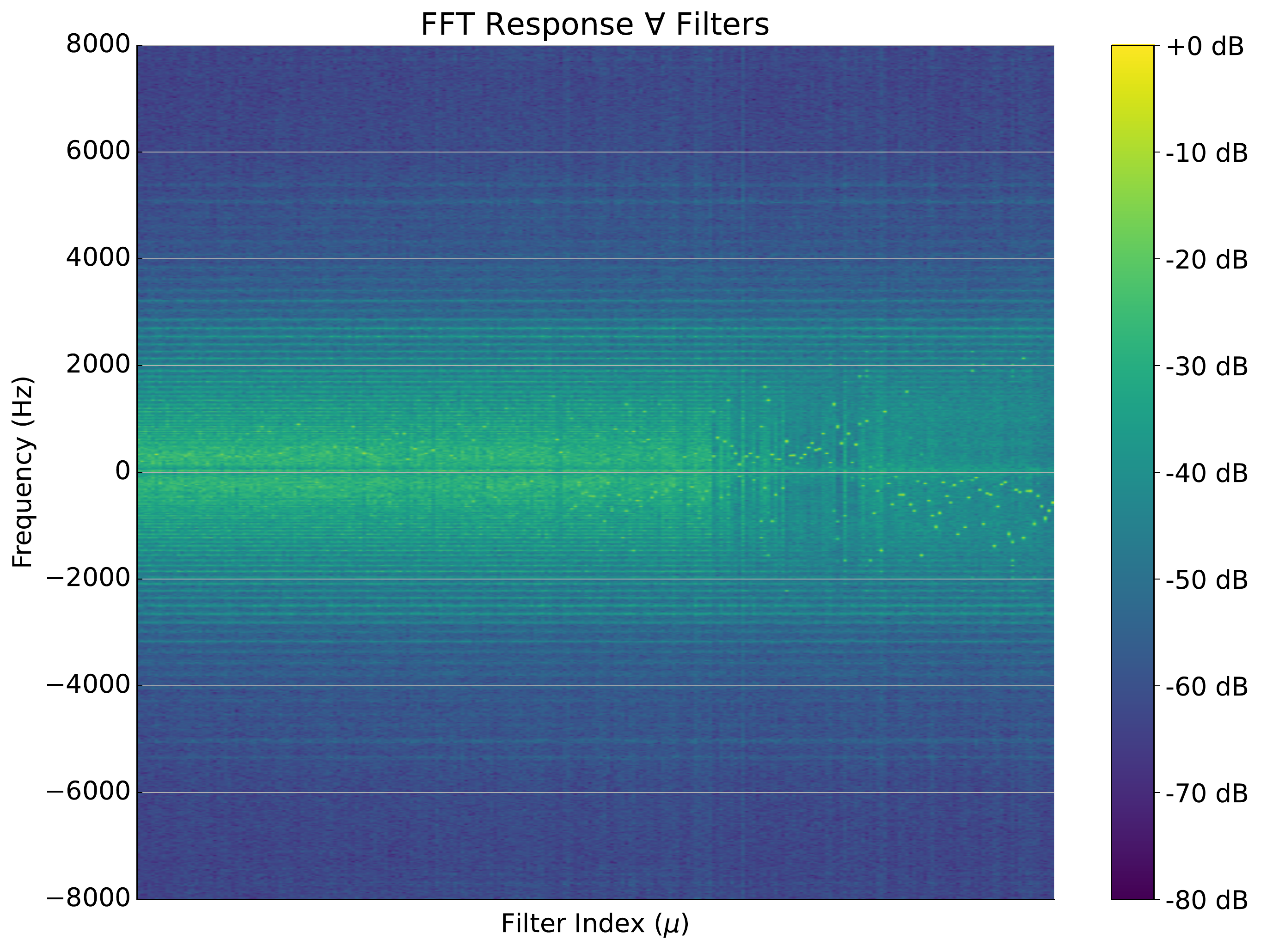}
\caption{Frequency response for all filters of the \textit{cl+rnd} experiment, ordered by spectral centroid.}
\end{figure}

\vfill

\begin{figure*}[!h]
\centerline{\includegraphics[width=1.05\linewidth]{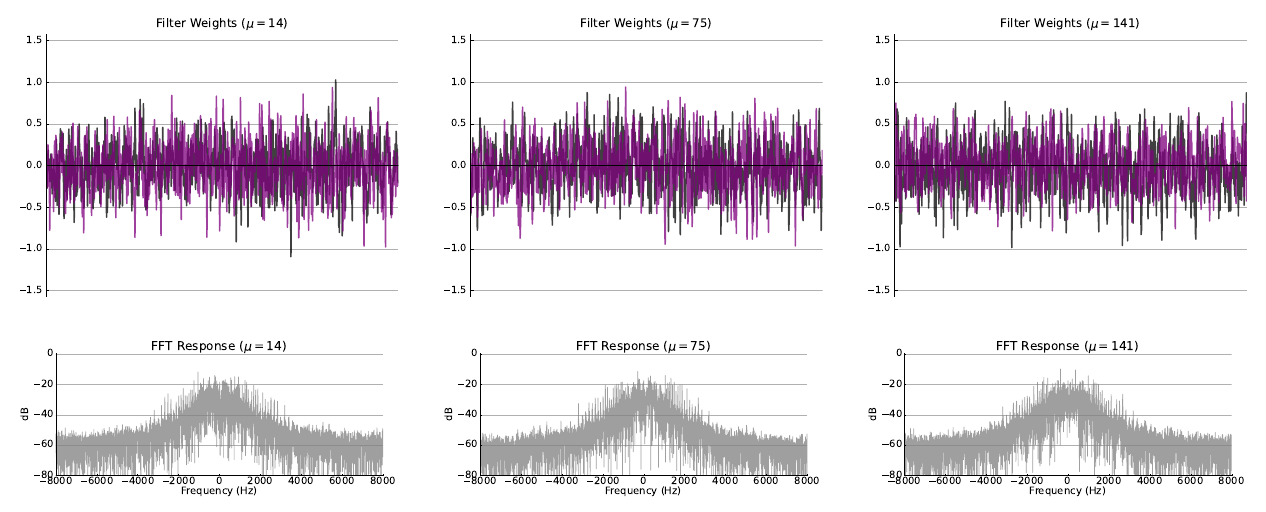}}
\caption{Examples of noisy filters from the \textit{cl+rnd} experiment.}
\end{figure*}

\begin{figure*}[!h]
\centerline{\includegraphics[width=1.05\linewidth]{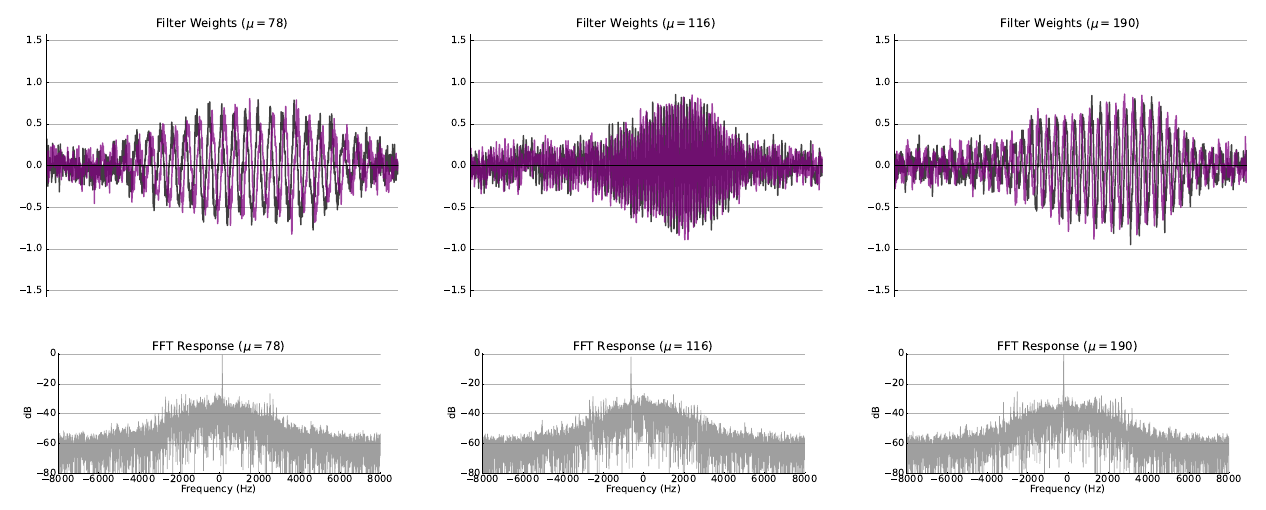}}
\caption{Examples of filters from the \textit{cl+rnd} experiment with a single prominent fundamental frequency.}
\end{figure*}

\begin{figure*}[!h]
\centerline{\includegraphics[width=1.05\linewidth]{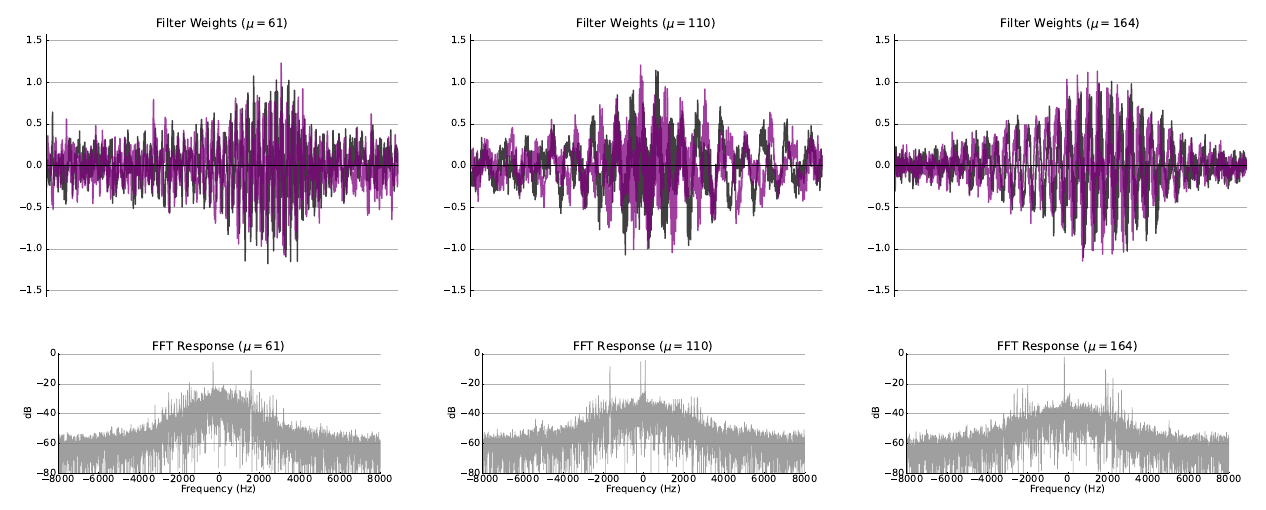}}
\caption{Examples of filters from the \textit{cl+rnd} experiment with multiple prominent fundamental frequencies.}
\end{figure*}

\begin{figure*}[!h]
\centerline{\includegraphics[width=1.05\linewidth]{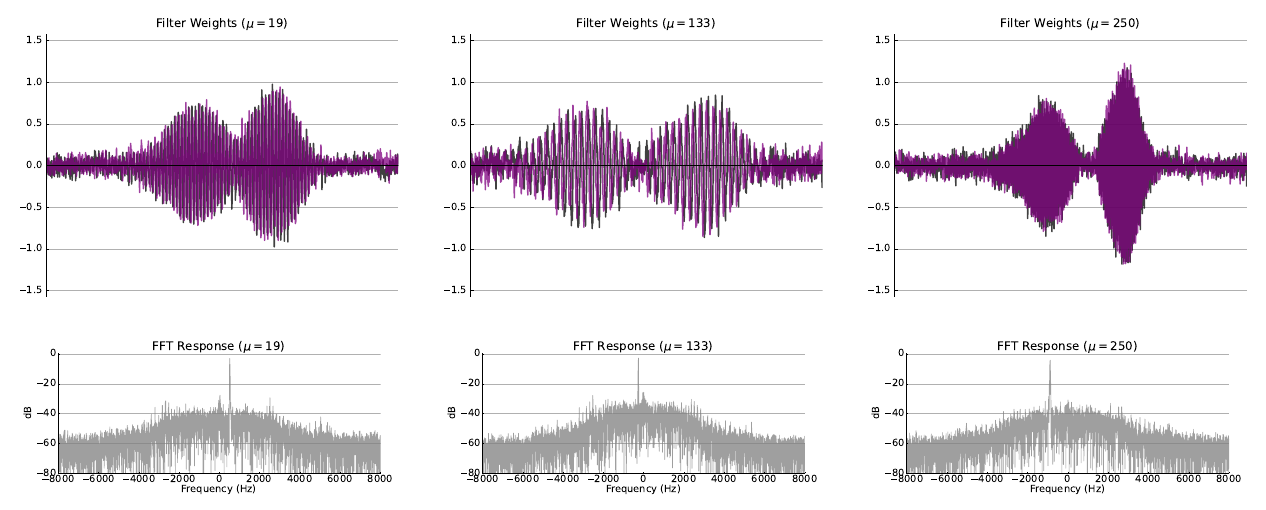}}
\caption{Examples of filters from the \textit{cl+rnd} experiment with two main lobes in the impulse response.}
\end{figure*}

\clearpage

\subsection{Classic + VQT (cl+vqt)}
\begin{figure}[!h]
\includegraphics[width=\textwidth]{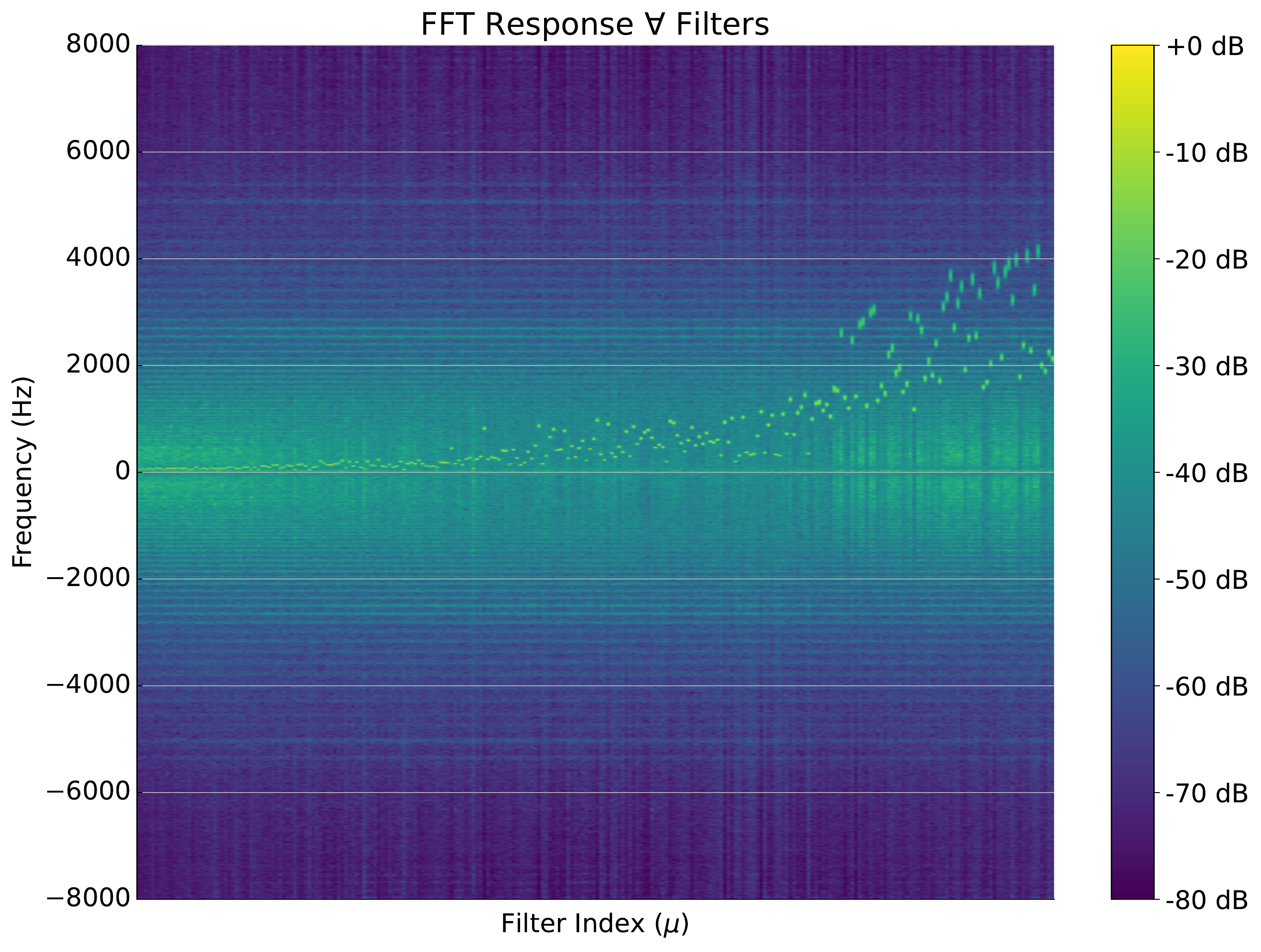}
\caption{Frequency response for all filters of the \textit{cl+vqt} experiment, ordered by spectral centroid.}
\label{fig:cl_vqt_filters}
\end{figure}

\vfill

\begin{figure*}[!h]
\centerline{\includegraphics[width=1.05\linewidth]{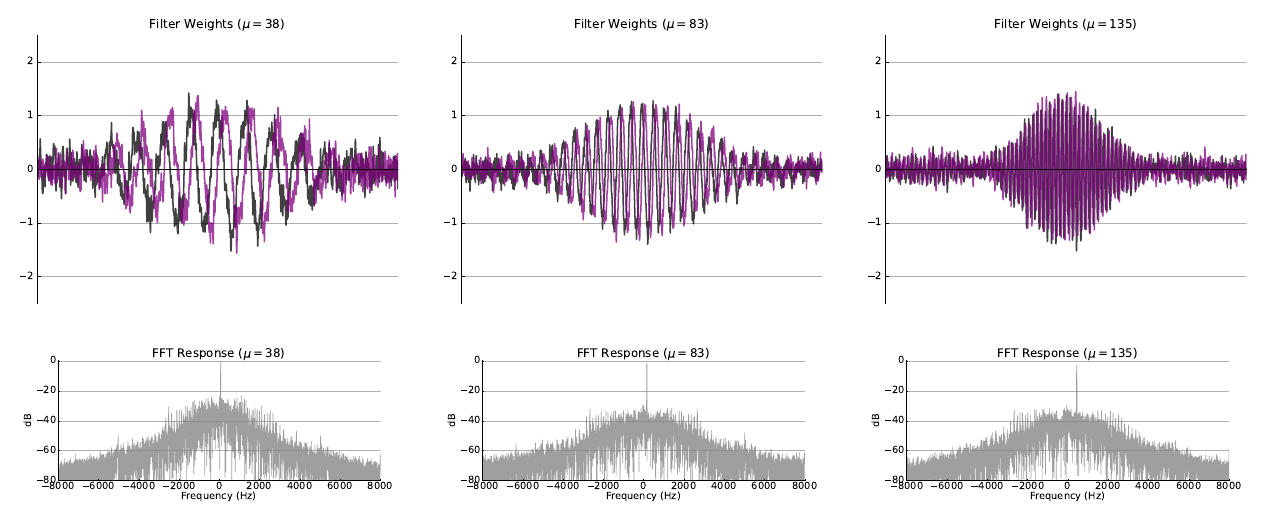}}
\caption{Examples of filters from the \textit{cl+vqt} experiment which remain mostly intact w.r.t. initialization.}
\end{figure*}

\begin{figure*}[!h]
\centerline{\includegraphics[width=1.04\linewidth]{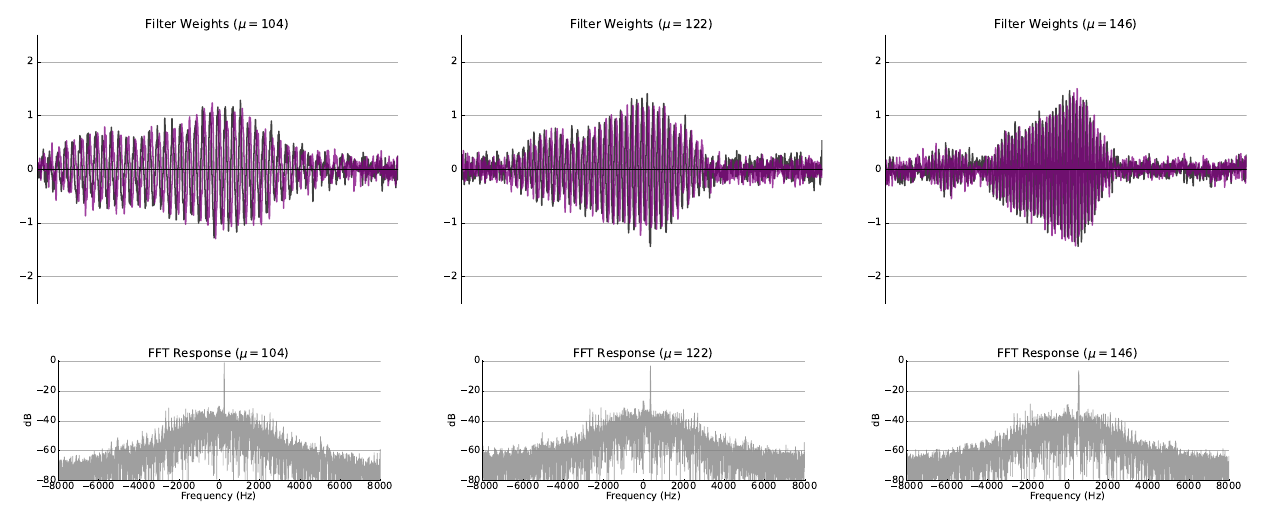}}
\caption{Examples of filters from the \textit{cl+vqt} experiment with impulse responses distorted w.r.t. initialization.}
\end{figure*}

\begin{figure*}[!h]
\centerline{\includegraphics[width=1.04\linewidth]{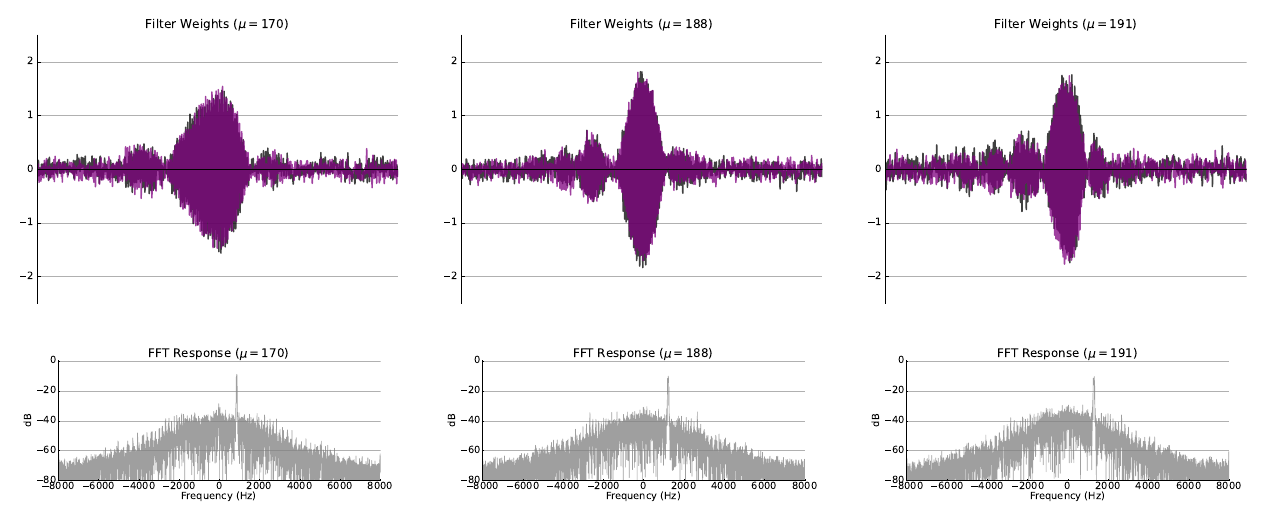}}
\caption{Examples of filters from the \textit{cl+vqt} experiment where side-lobes were introduced to the impulse response.}
\end{figure*}

\begin{figure*}[!h]
\centerline{\includegraphics[width=1.04\linewidth]{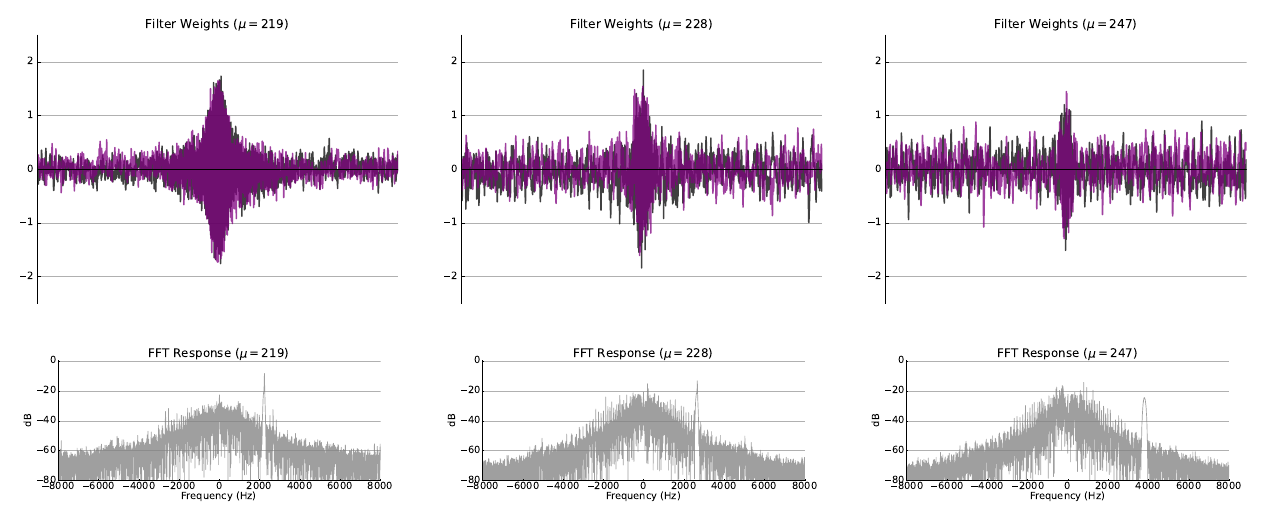}}
\caption{Examples of high-frequency filters from the \textit{cl+vqt} experiment.}
\end{figure*}

\clearpage

\subsection{Hilbert + Random (hb+rnd)}
\begin{figure}[!h]
\includegraphics[width=\textwidth]{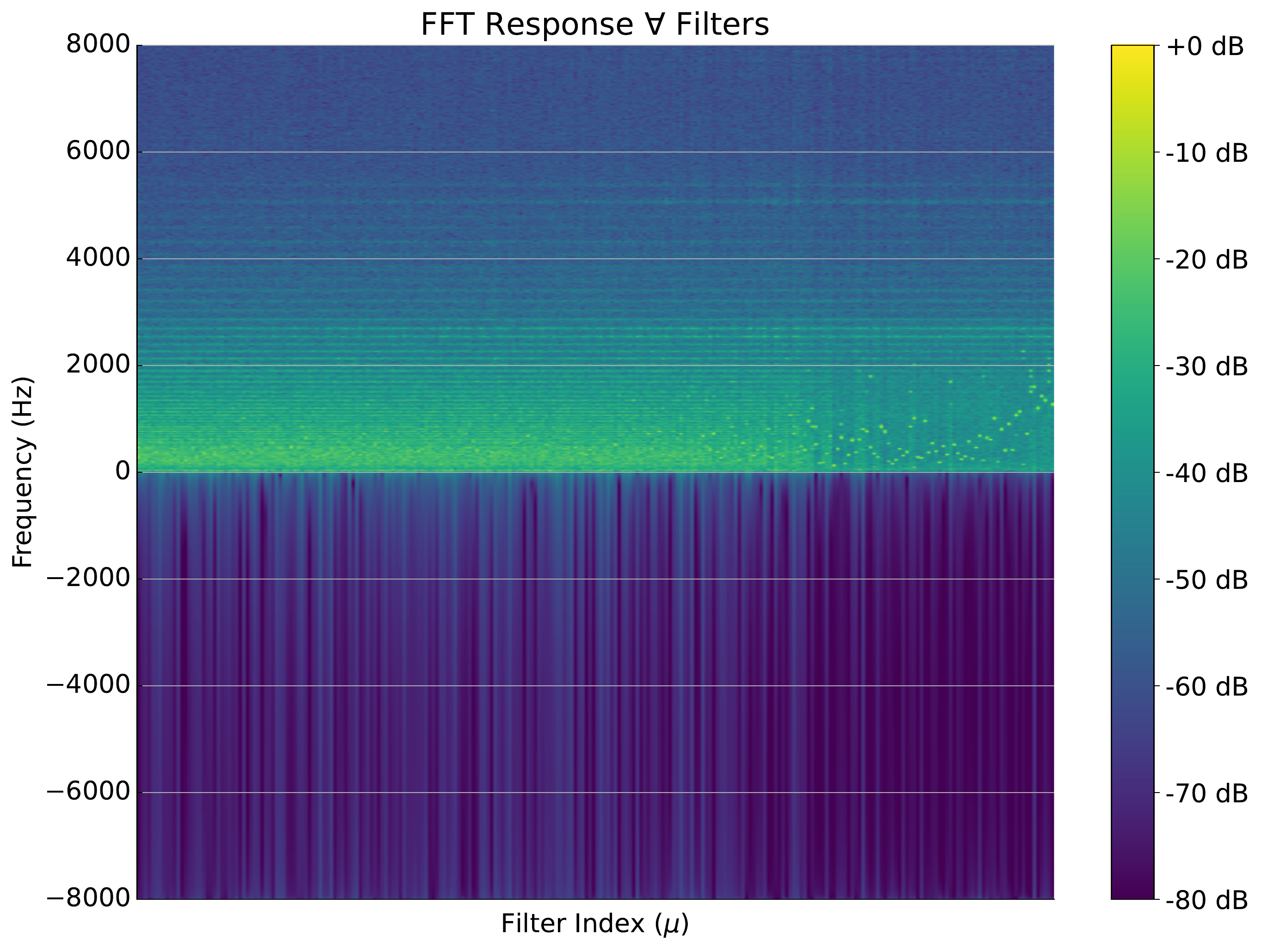}
\caption{Frequency response for all filters of the \textit{hb+rnd} experiment, ordered by spectral centroid.}
\label{fig:hb_rnd_filters}
\end{figure}

\vfill

\begin{figure*}[!h]
\centerline{\includegraphics[width=1.05\linewidth]{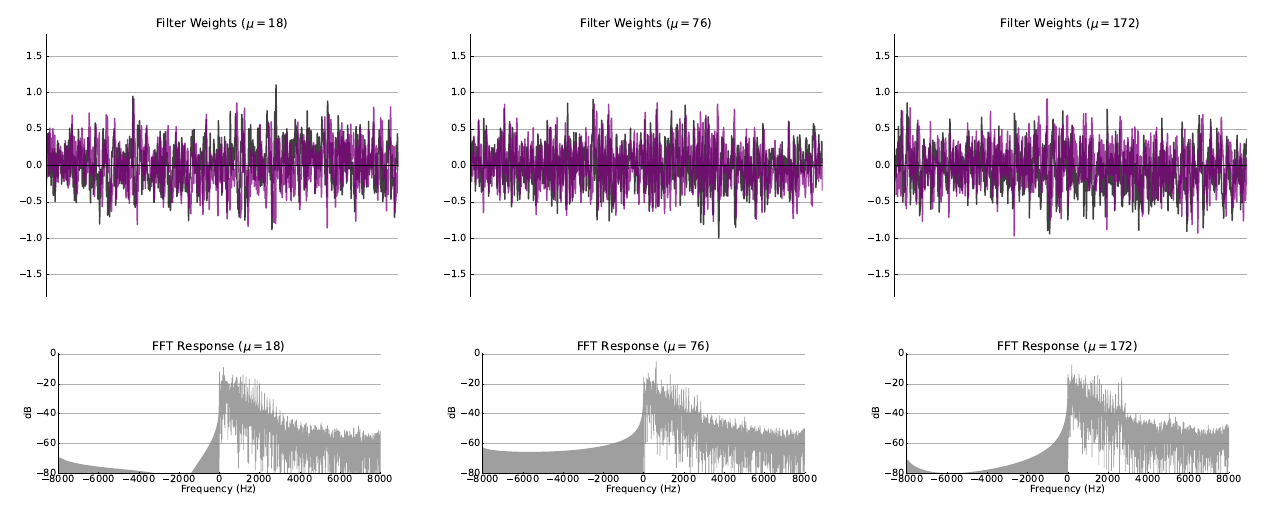}}
\caption{Examples of noisy filters from the \textit{hb+rnd} experiment.}
\end{figure*}

\begin{figure*}[!h]
\centerline{\includegraphics[width=1.05\linewidth]{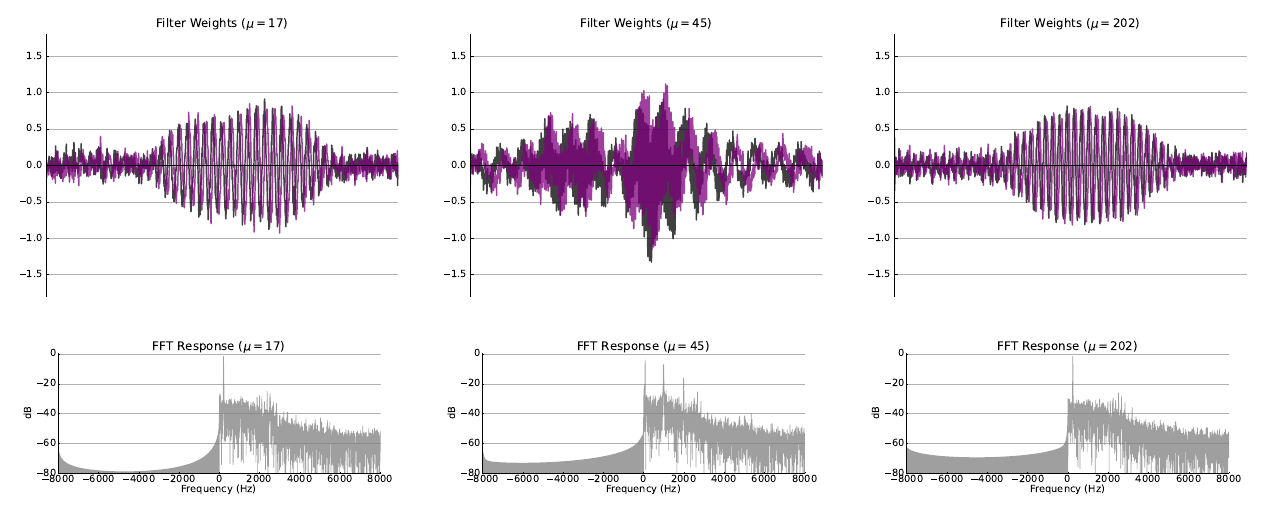}}
\caption{Examples of filters from the \textit{hb+rnd} experiment with a single prominent fundamental frequency.}
\end{figure*}

\begin{figure*}[!h]
\centerline{\includegraphics[width=1.05\linewidth]{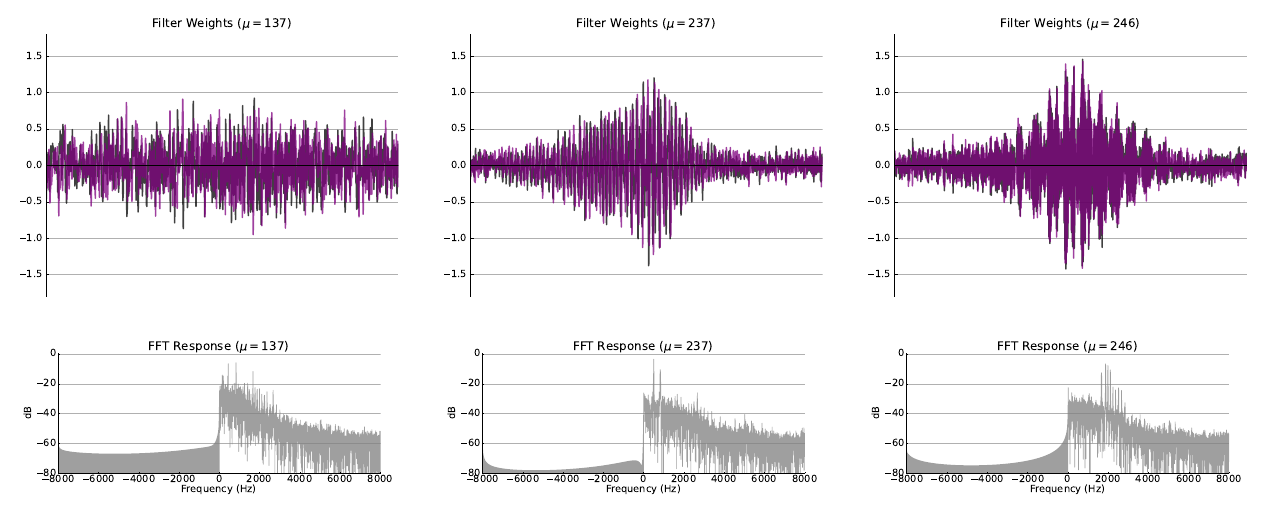}}
\caption{Examples of filters from the \textit{hb+rnd} experiment with multiple prominent fundamental frequencies.}
\end{figure*}

\begin{figure*}[!h]
\centerline{\includegraphics[width=1.05\linewidth]{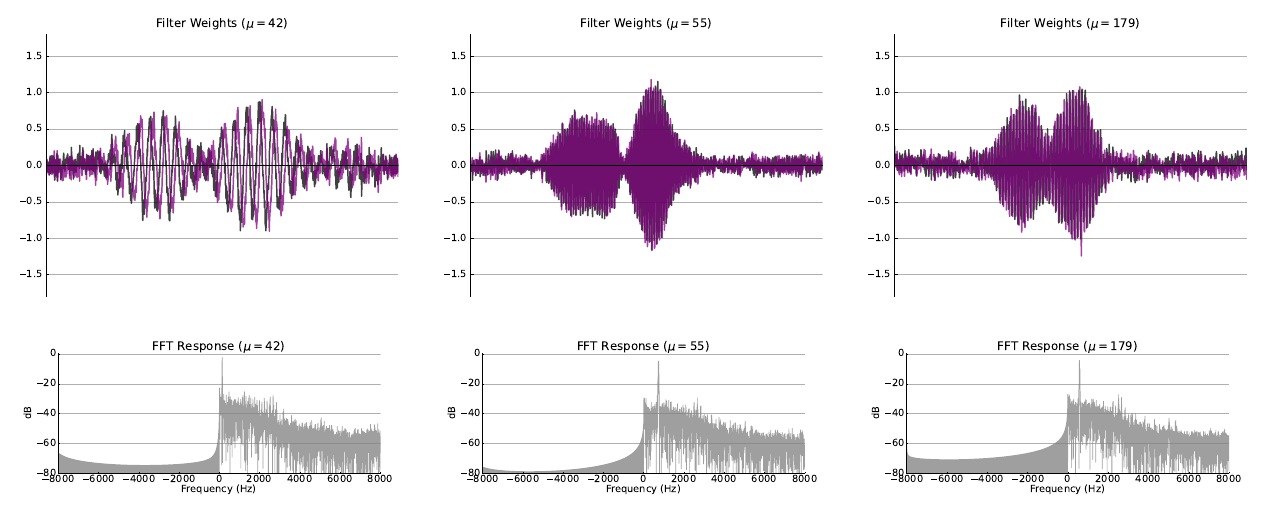}}
\caption{Examples of filters from the \textit{hb+rnd} experiment with two main lobes in the impulse response.}
\end{figure*}

\clearpage

\subsection{Hilbert + VQT (hb+vqt)}
\begin{figure}[!h]
\includegraphics[width=\textwidth]{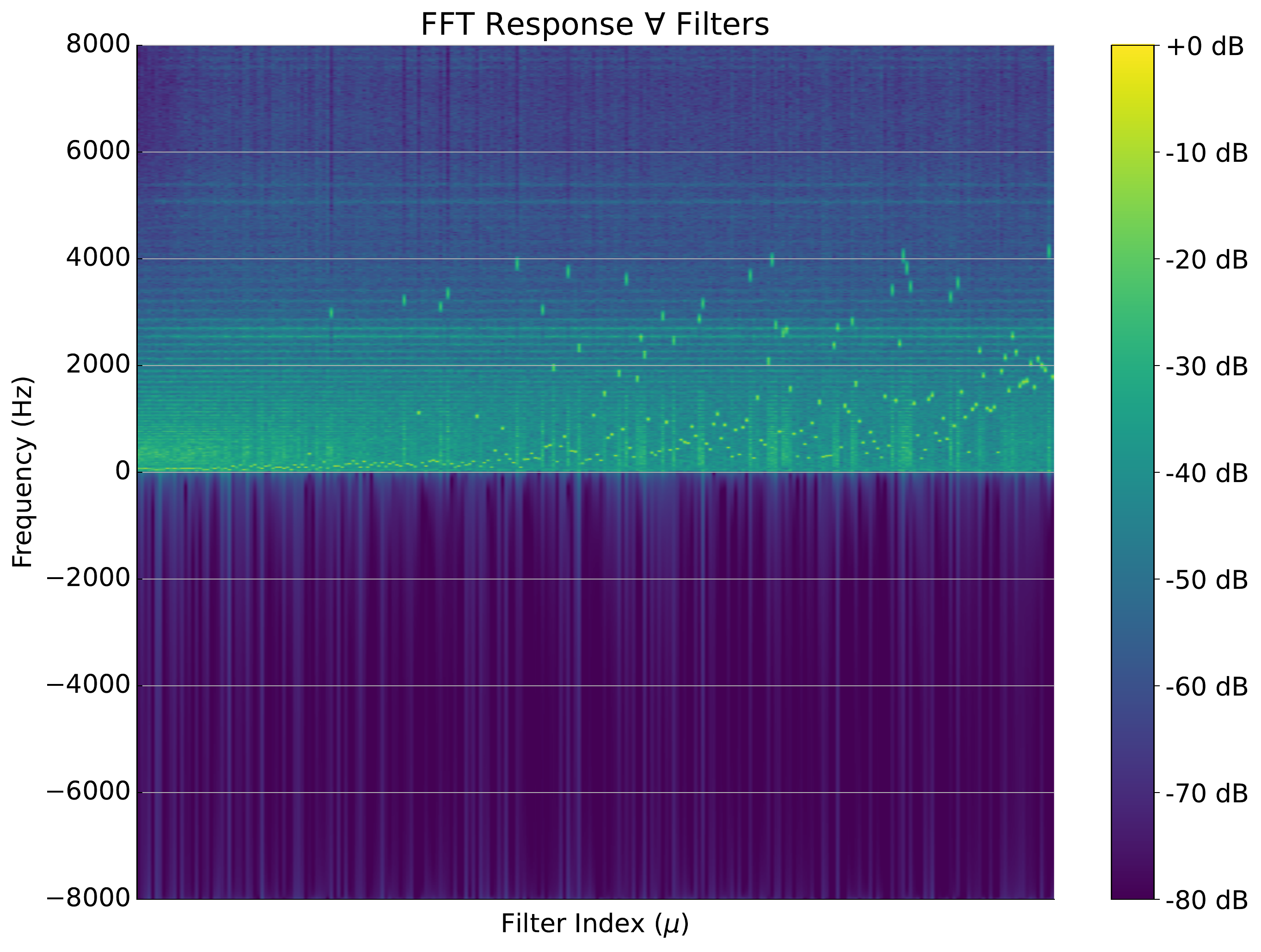}
\caption{Frequency response for all filters of the \textit{hb+vqt} experiment, ordered by spectral centroid.}
\label{fig:hb_vqt_filters}
\end{figure}

\vfill

\begin{figure*}[!h]
\centerline{\includegraphics[width=1.05\linewidth]{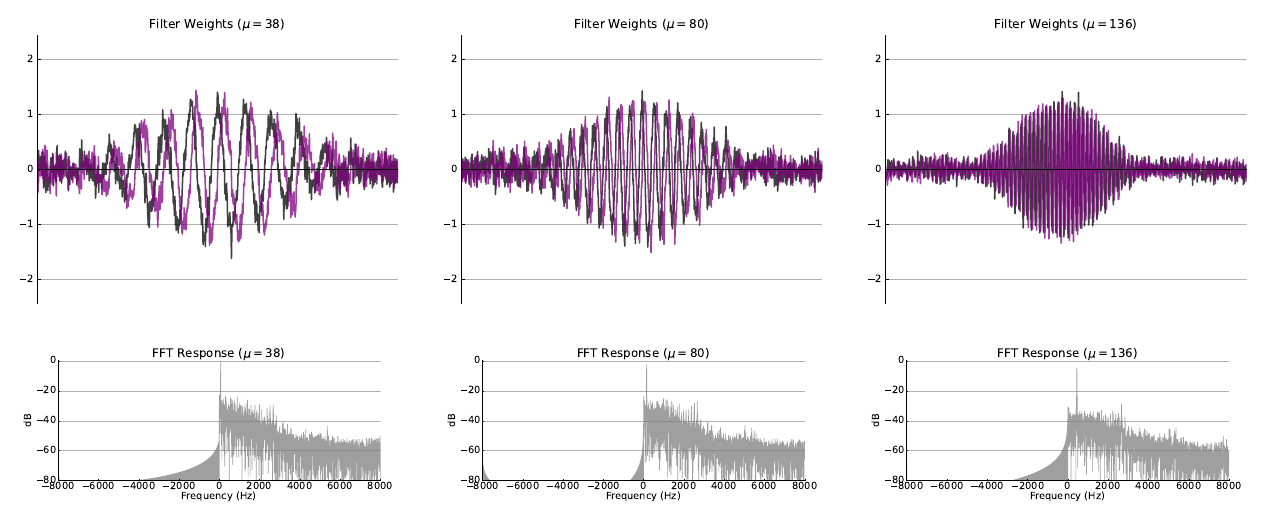}}
\caption{Examples of filters from the \textit{hb+vqt} experiment which remain mostly intact w.r.t. initialization.}
\end{figure*}

\begin{figure*}[!h]
\centerline{\includegraphics[width=1.04\linewidth]{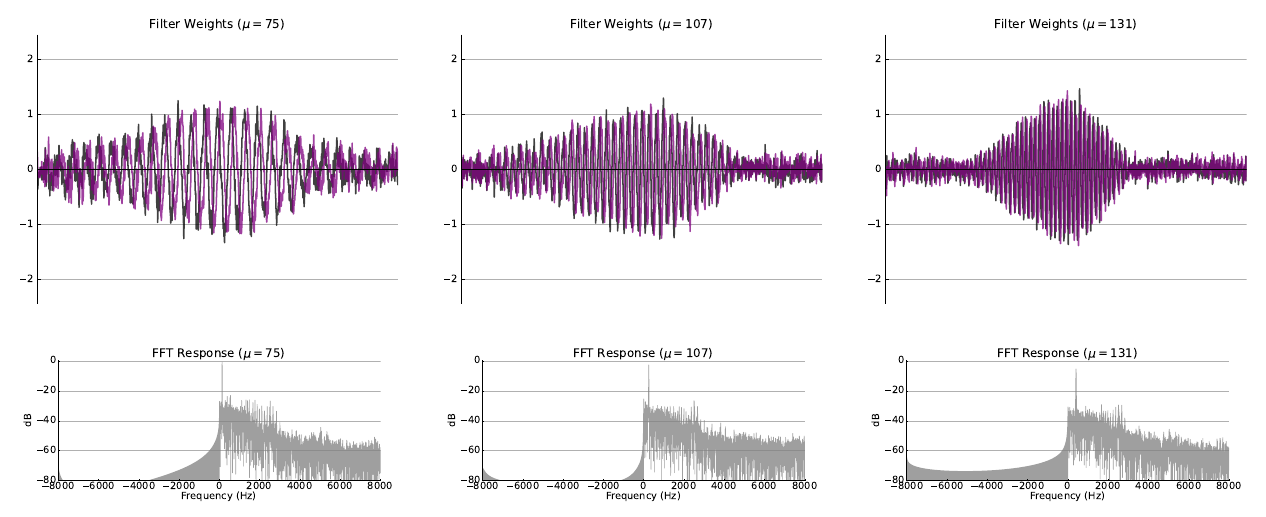}}
\caption{Examples of filters from the \textit{hb+vqt} experiment with impulse responses distorted w.r.t. initialization.}
\end{figure*}

\begin{figure*}[!h]
\centerline{\includegraphics[width=1.04\linewidth]{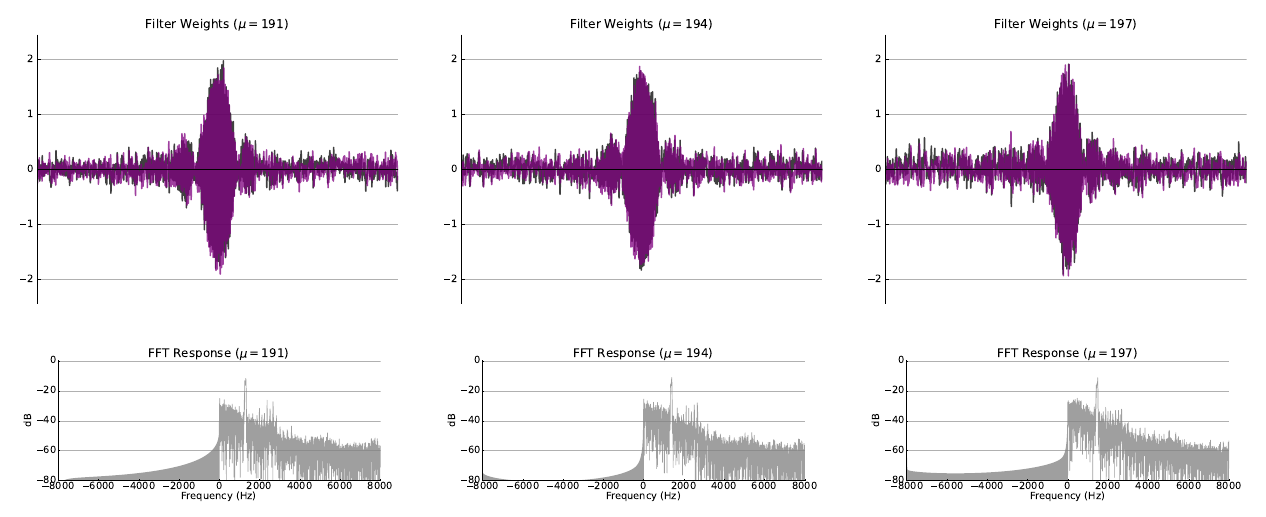}}
\caption{Examples of filters from the \textit{hb+vqt} experiment where side-lobes were introduced to the impulse response.}
\end{figure*}

\begin{figure*}[!h]
\centerline{\includegraphics[width=1.04\linewidth]{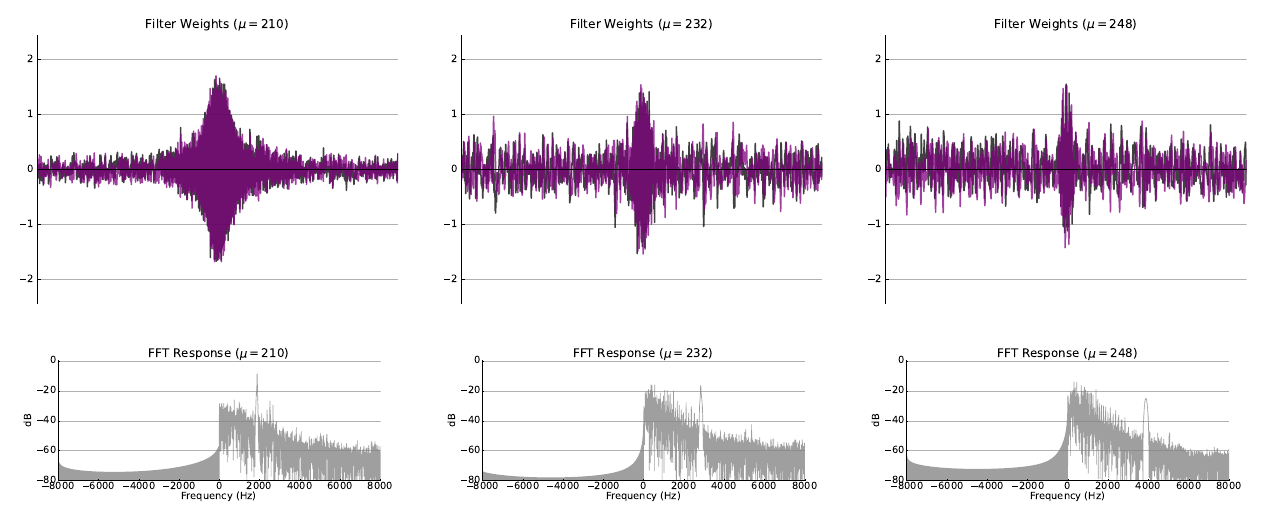}}
\caption{Examples of high-frequency filters from the \textit{hb+vqt} experiment.}
\end{figure*}

\clearpage

\subsection{Hilbert + Random + Bernoulli (hb+rnd+brn)}
\begin{figure}[!h]
\includegraphics[width=\textwidth]{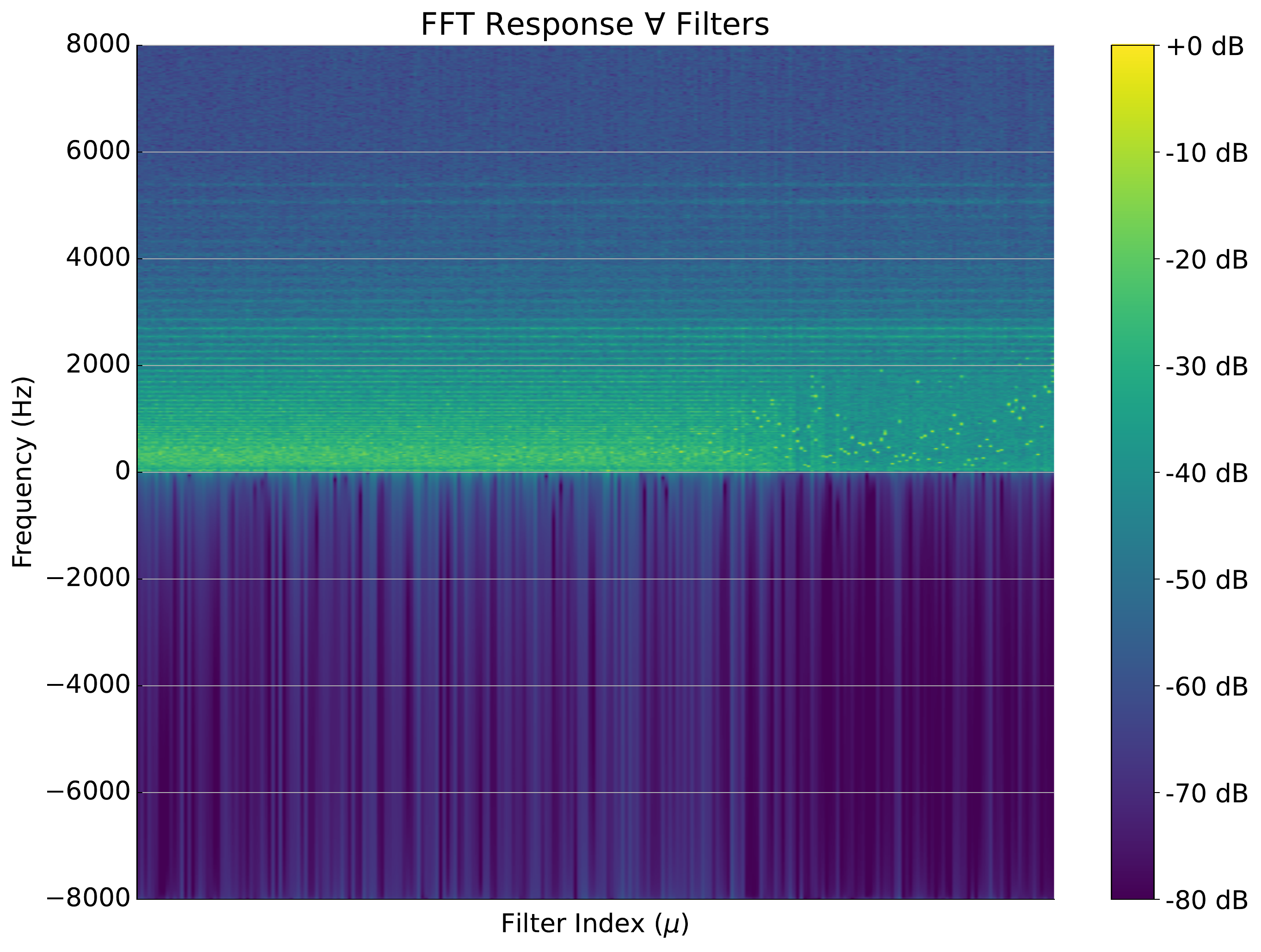}
\caption{Frequency response for all filters of the \textit{hb+rnd+brn} experiment, ordered by spectral centroid.}
\label{fig:hb_rnd_brn_filters}
\end{figure}

\vfill

\begin{figure*}[!h]
\centerline{\includegraphics[width=1.05\linewidth]{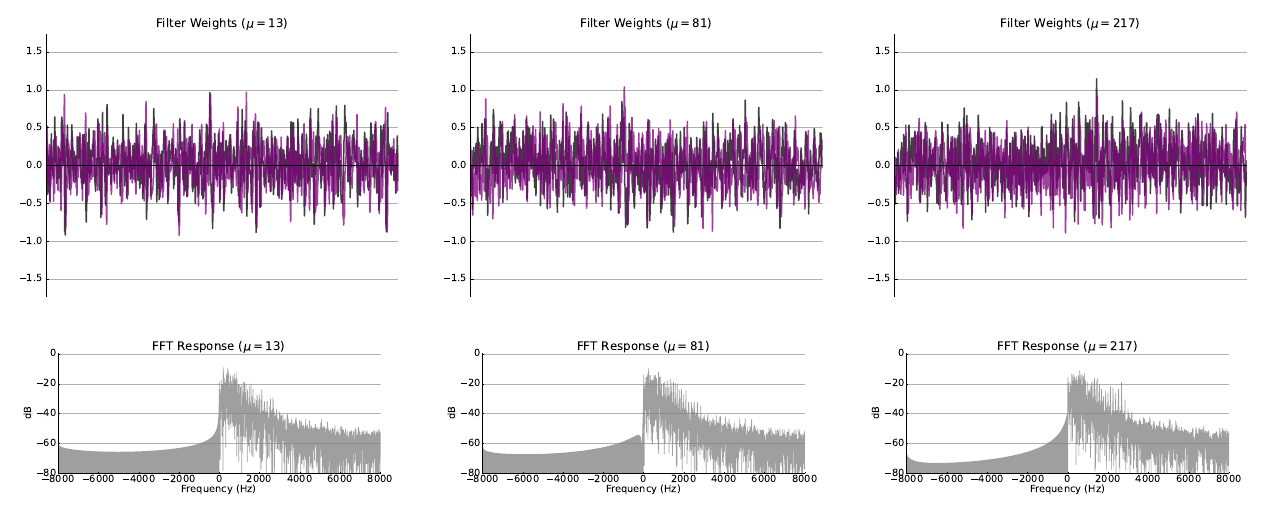}}
\caption{Examples of noisy filters from the \textit{hb+rnd+brn} experiment.}
\end{figure*}

\begin{figure*}[!h]
\centerline{\includegraphics[width=1.05\linewidth]{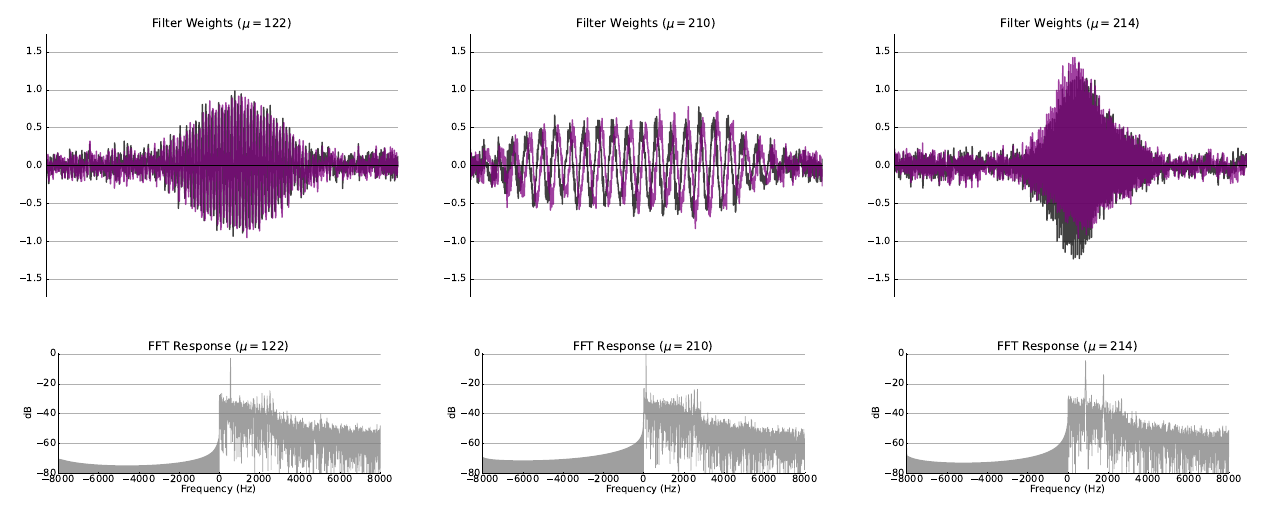}}
\caption{Examples of filters from the \textit{hb+rnd+brn} experiment with a single prominent fundamental frequency.}
\end{figure*}

\begin{figure*}[!h]
\centerline{\includegraphics[width=1.05\linewidth]{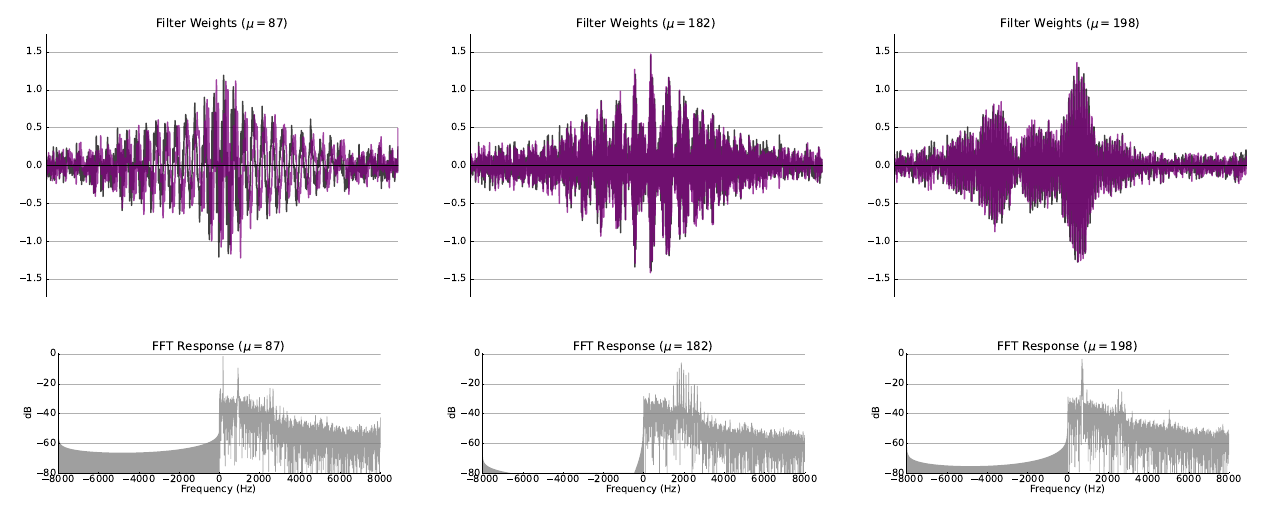}}
\caption{Examples of filters from the \textit{hb+rnd+brn} experiment with multiple prominent fundamental frequencies.}
\end{figure*}

\begin{figure*}[!h]
\centerline{\includegraphics[width=1.05\linewidth]{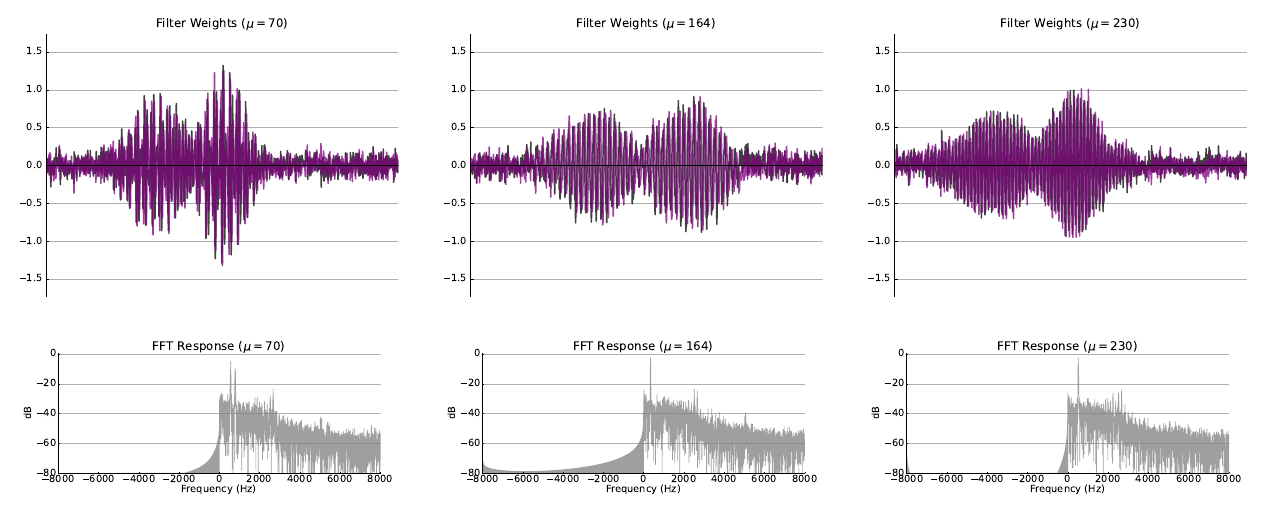}}
\caption{Examples of filters from the \textit{hb+rnd+brn} experiment with two main lobes in the impulse response.}
\end{figure*}

\clearpage

\subsection{Hilbert + Random + Gaussian (hb+rnd+gau)}
\begin{figure}[!h]
\includegraphics[width=\textwidth]{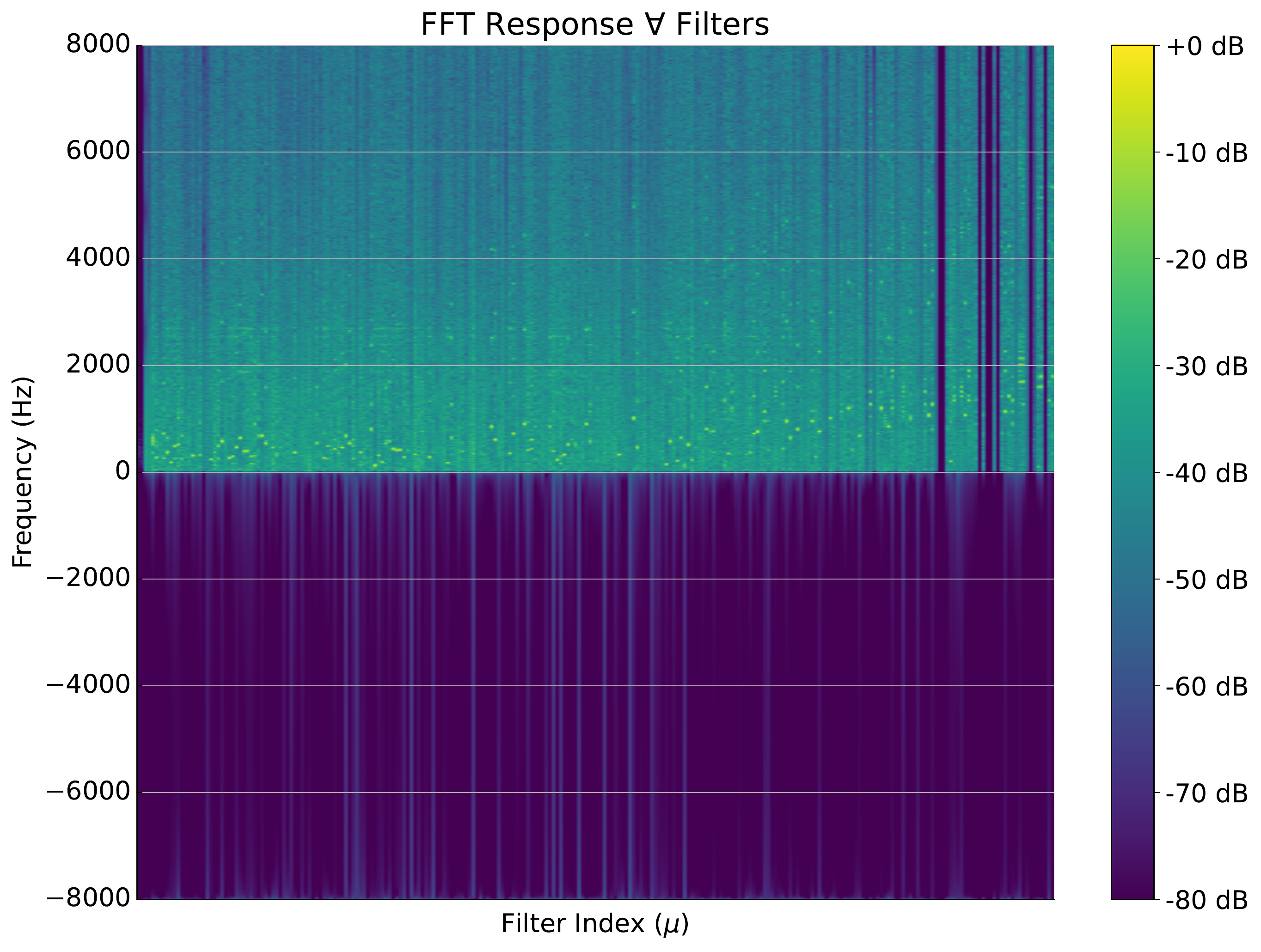}
\caption{Frequency response for all filters of the \textit{hb+rnd+gau} experiment, ordered by spectral centroid.}
\label{fig:hb_rnd_gau_filters}
\end{figure}

\vfill

\begin{figure*}[!h]
\centerline{\includegraphics[width=1.05\linewidth]{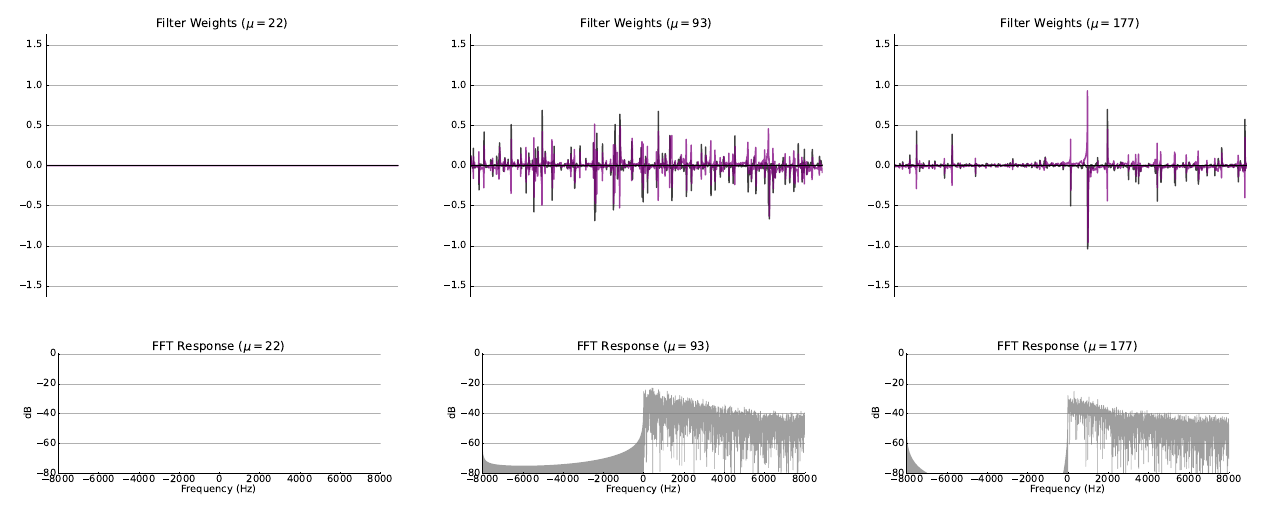}}
\caption{Examples of sparse filters from the \textit{hb+rnd+gau} experiment.}
\end{figure*}

\begin{figure*}[!h]
\centerline{\includegraphics[width=1.05\linewidth]{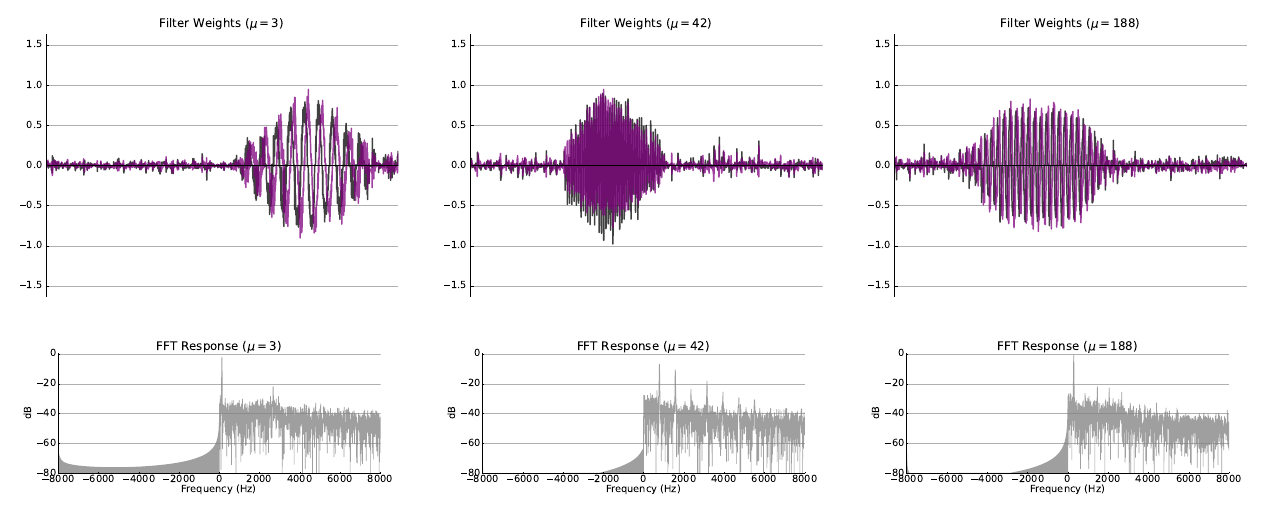}}
\caption{Examples of filters from the \textit{hb+rnd+gau} experiment with a single prominent fundamental frequency.}
\end{figure*}

\begin{figure*}[!h]
\centerline{\includegraphics[width=1.05\linewidth]{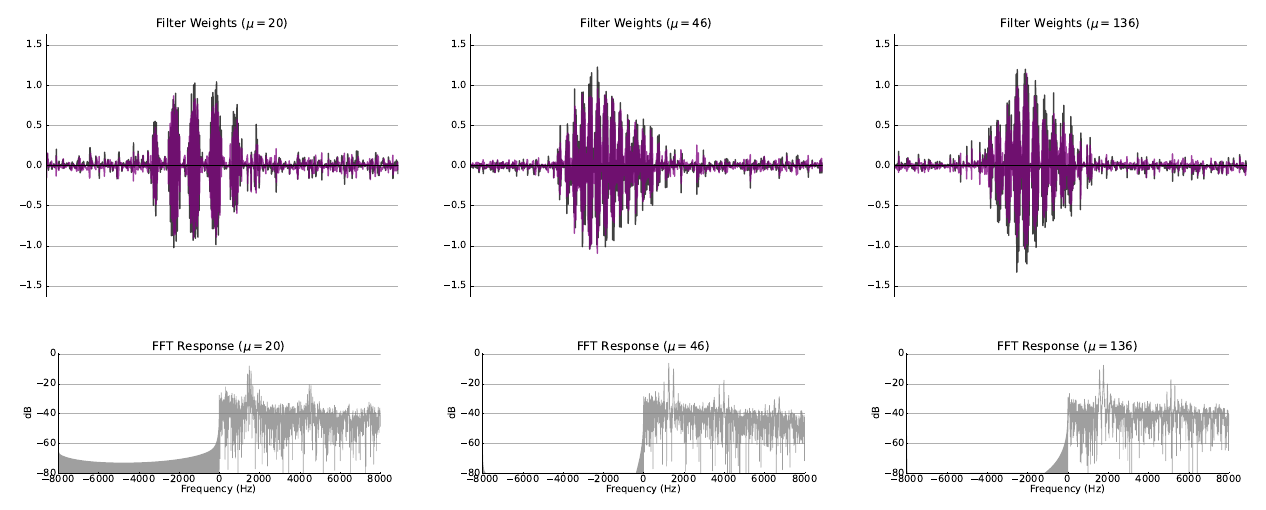}}
\caption{Examples of filters from the \textit{hb+rnd+gau} experiment with multiple prominent fundamental frequencies.}
\end{figure*}

\begin{figure*}[!h]
\centerline{\includegraphics[width=1.05\linewidth]{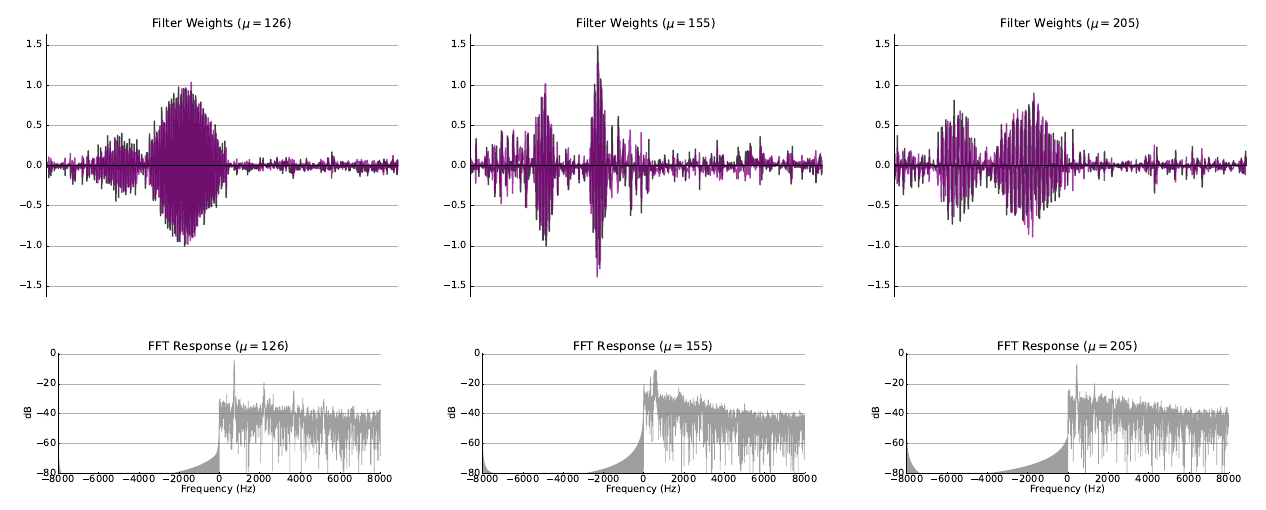}}
\caption{Examples of filters from the \textit{hb+rnd+gau} experiment with two main lobes in the impulse response.}
\end{figure*}

\clearpage

\subsection{Hilbert + VQT + Variational (hb+vqt+var)}
\begin{figure}[!h]
\includegraphics[width=\textwidth]{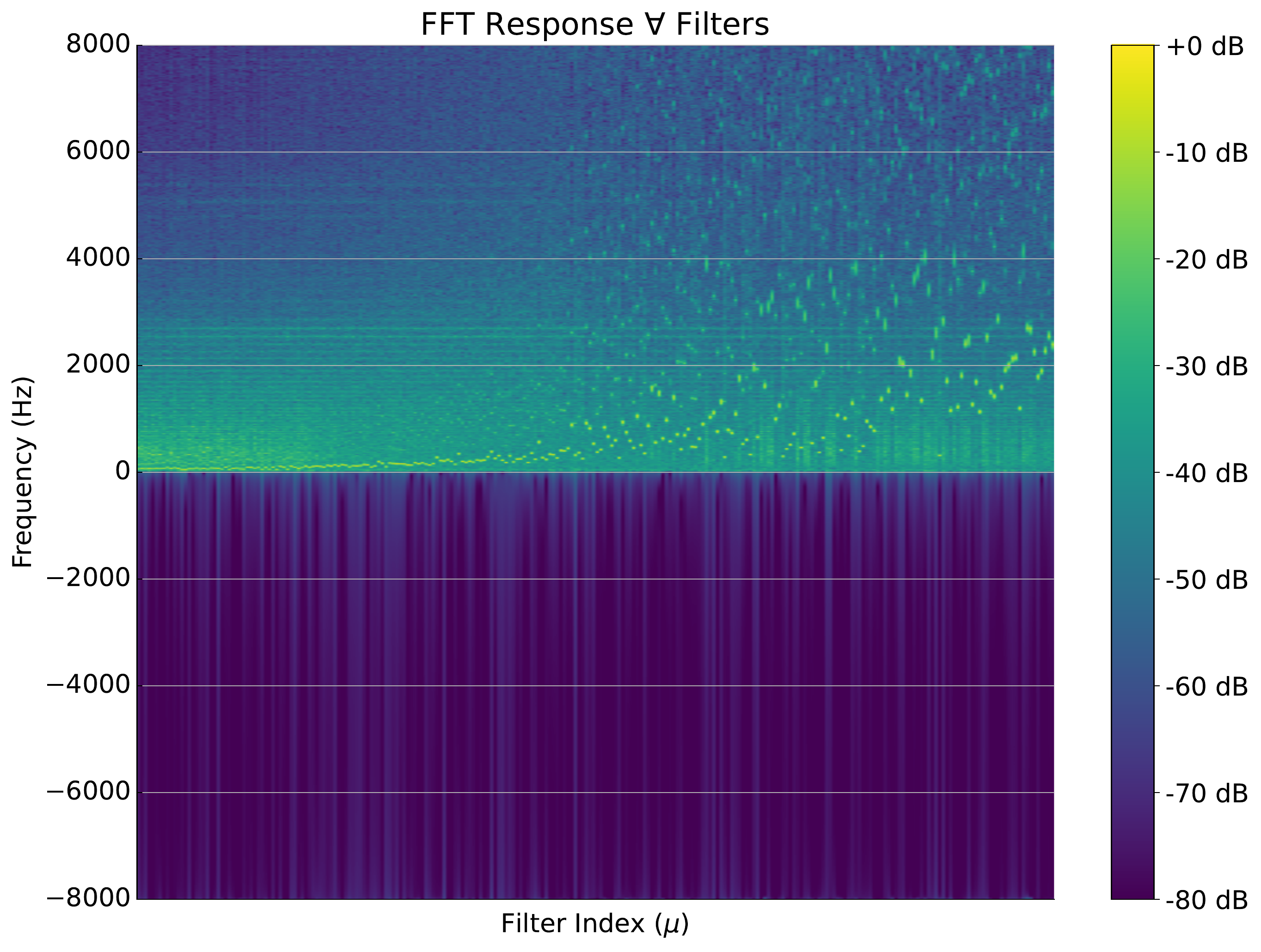}
\caption{Frequency response for all filters of the \textit{hb+vqt+var} experiment, ordered by spectral centroid.}
\label{fig:hb_vqt_var_filters}
\end{figure}

\vfill

\begin{figure*}[!h]
\centerline{\includegraphics[width=1.05\linewidth]{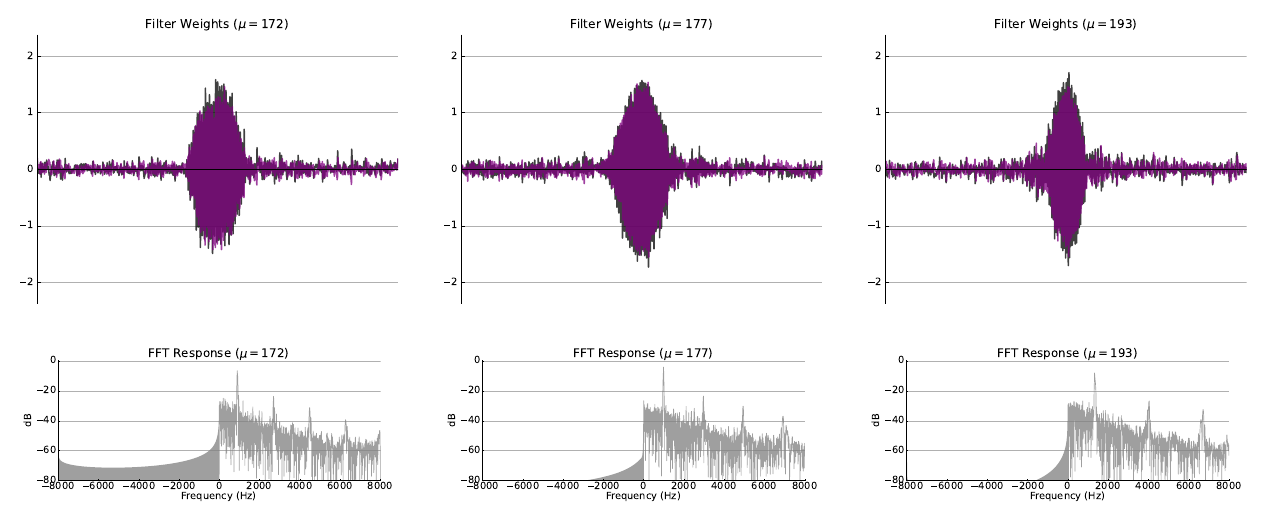}}
\caption{Examples of filters from the \textit{hb+vqt+var} experiment where harmonics were introduced.}
\end{figure*}

\begin{figure*}[!h]
\centerline{\includegraphics[width=1.05\linewidth]{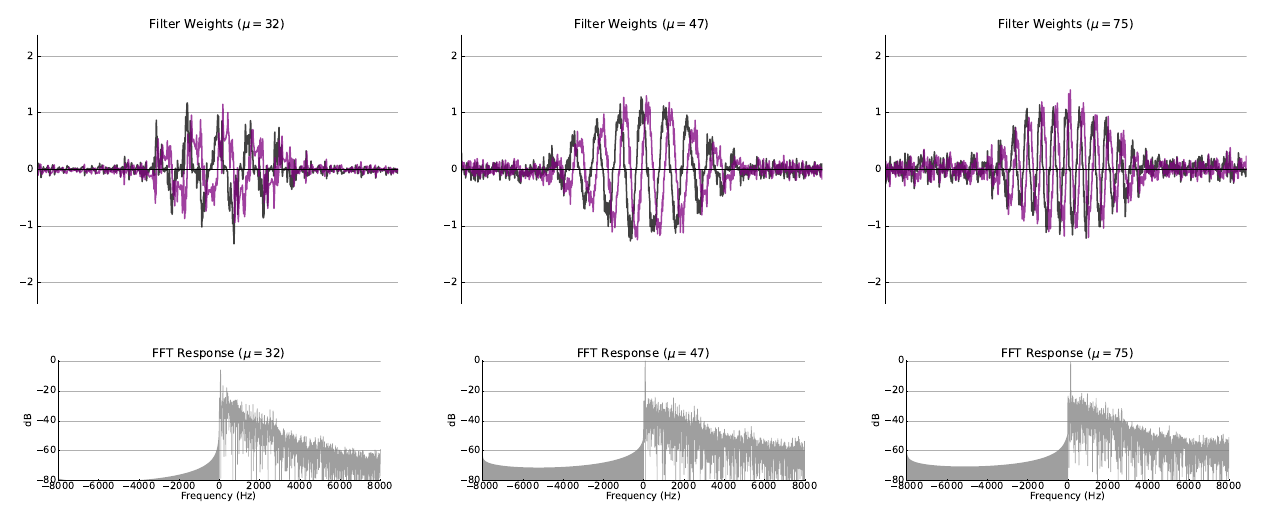}}
\centerline{\includegraphics[width=1.05\linewidth]{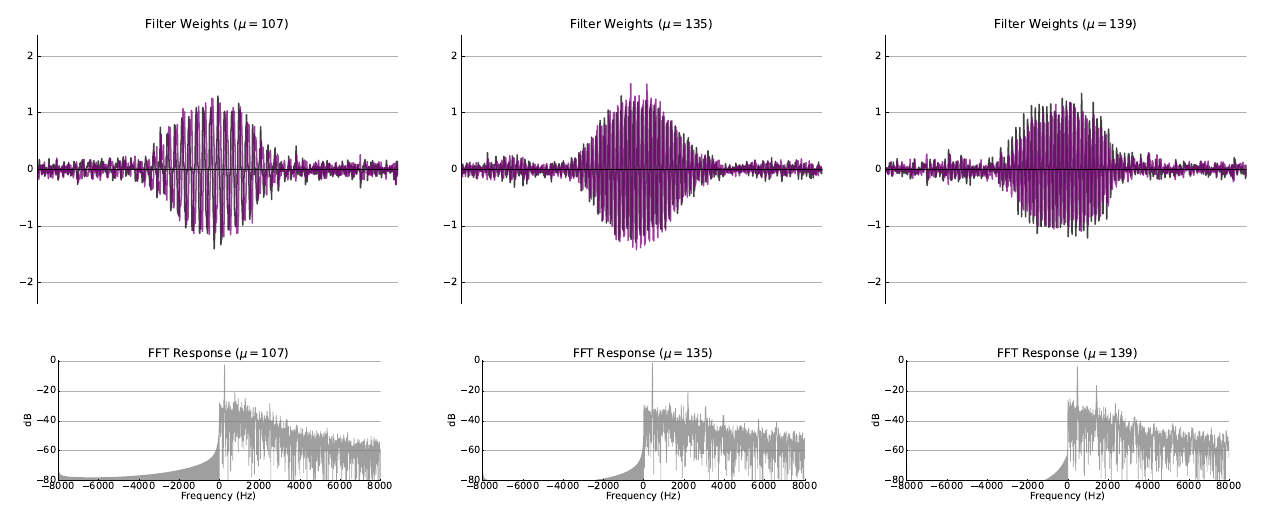}}
\caption{Examples of low- and mid-frequency filters from the \textit{hb+vqt+var} experiment.}
\end{figure*}

\vfill

\begin{figure*}[!h]
\centerline{\includegraphics[width=1.05\linewidth]{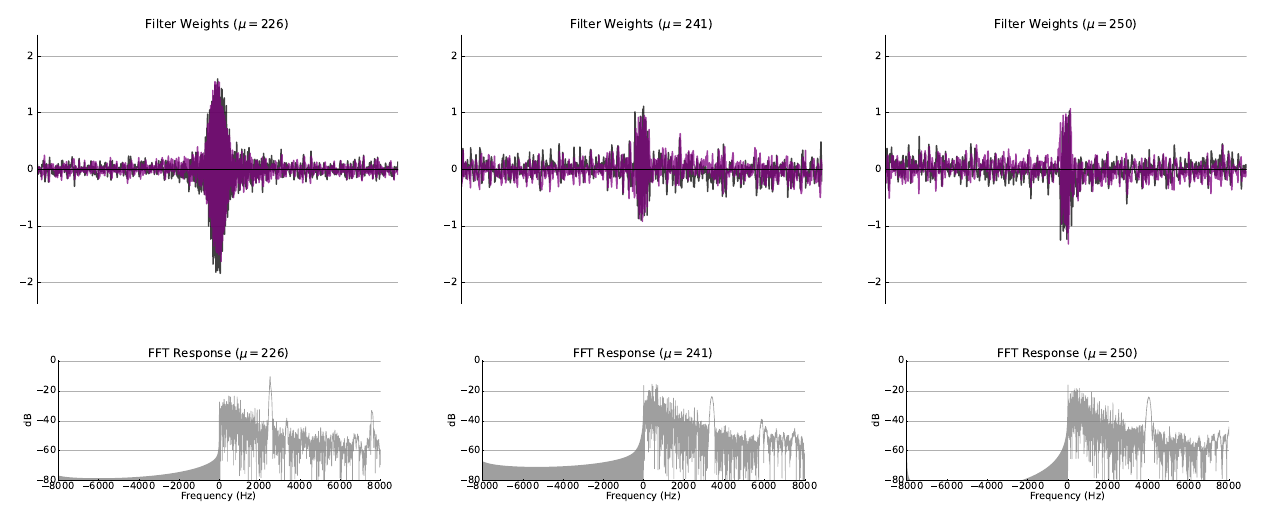}}
\caption{Examples of high-frequency filters from the \textit{hb+vqt+var} experiment.}
\end{figure*}

\clearpage

\subsection{Hilbert + Comb + Variational (hb+comb+var)}
\begin{figure}[!h]
\includegraphics[width=\textwidth]{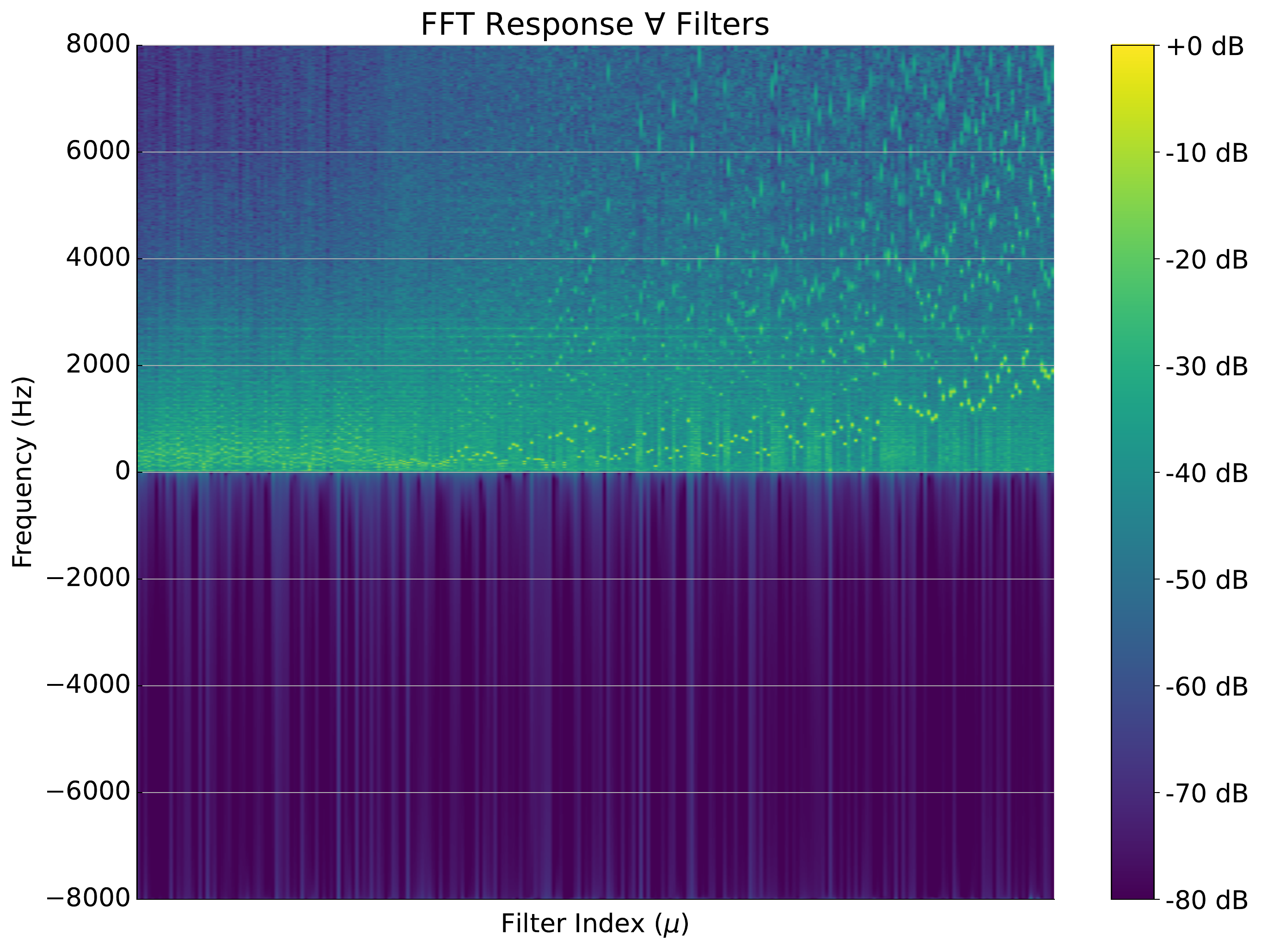}
\caption{Frequency response for all filters of the \textit{hb+comb+var} experiment, ordered by spectral centroid.}
\label{fig:hb_comb_var_filters}
\end{figure}

\vfill

\begin{figure*}[!h]
\centerline{\includegraphics[width=1.05\linewidth]{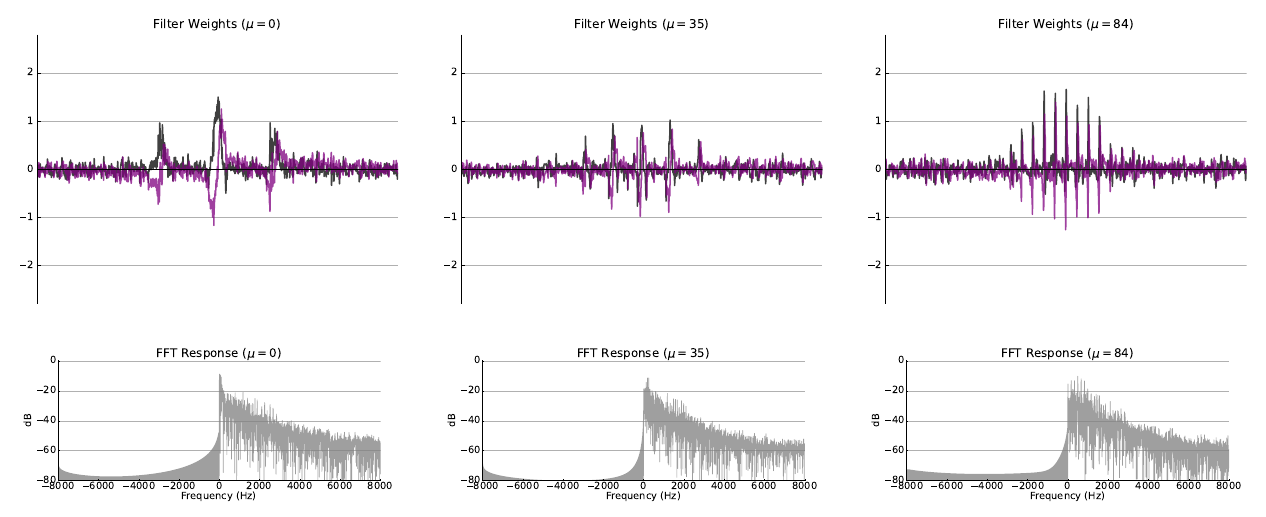}}
\caption{Examples of filters from the \textit{hb+comb+var} experiment which retain most of the comb initialization. }
\end{figure*}

\begin{figure*}[!h]
\centerline{\includegraphics[width=1.04\linewidth]{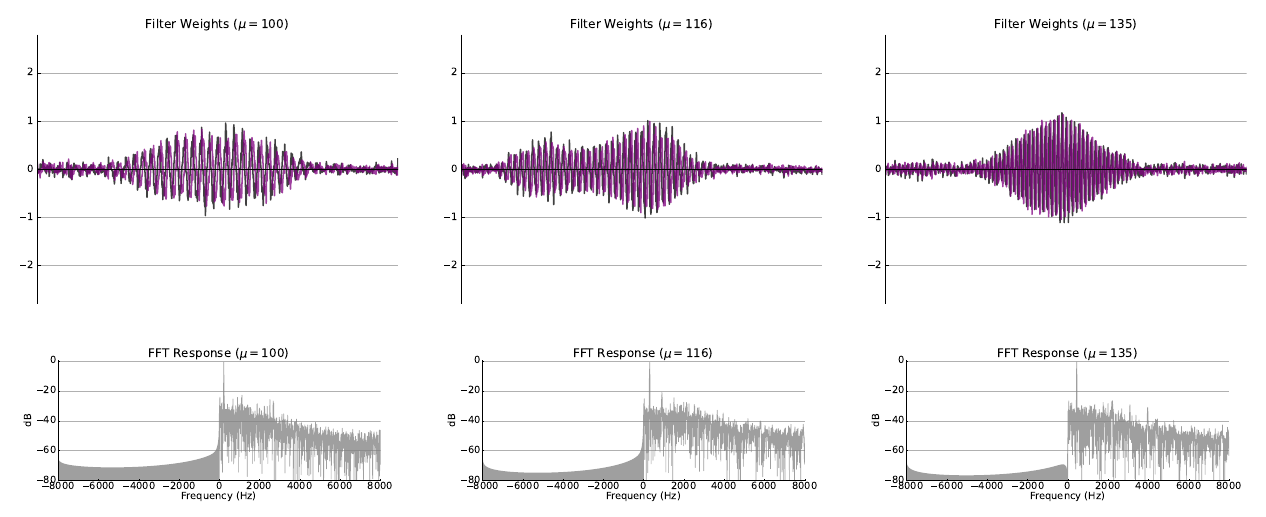}}
\caption{Examples of filters from the \textit{hb+comb+var} experiment with a single prominent fundamental frequency.}
\end{figure*}

\begin{figure*}[!h]
\centerline{\includegraphics[width=1.04\linewidth]{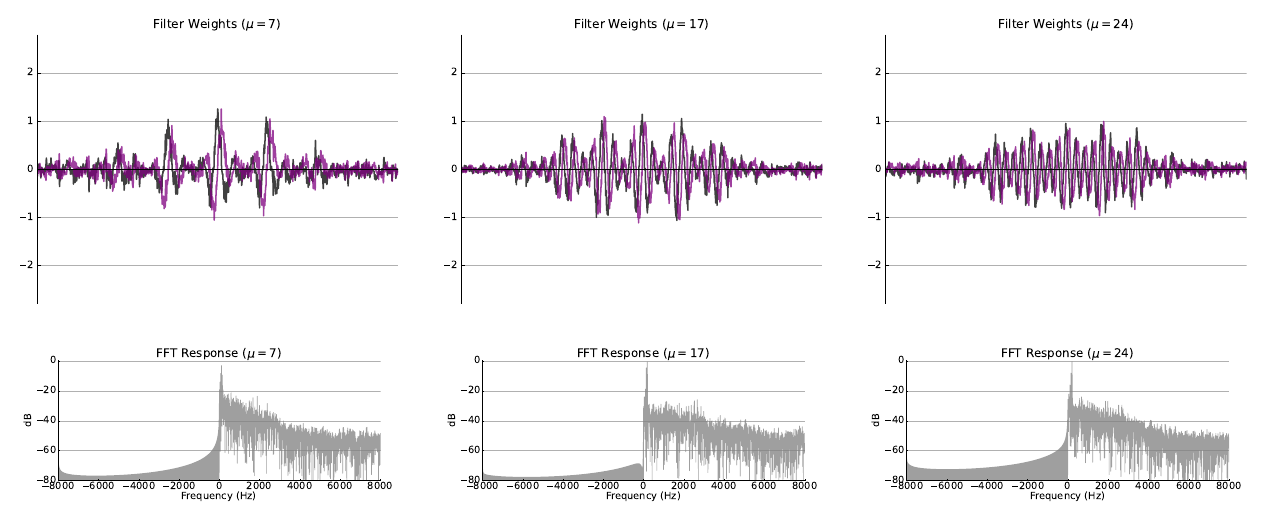}}
\caption{Examples of filters from the \textit{hb+comb+var} experiment with a comb-modulated response.}
\end{figure*}

\begin{figure*}[!h]
\centerline{\includegraphics[width=1.04\linewidth]{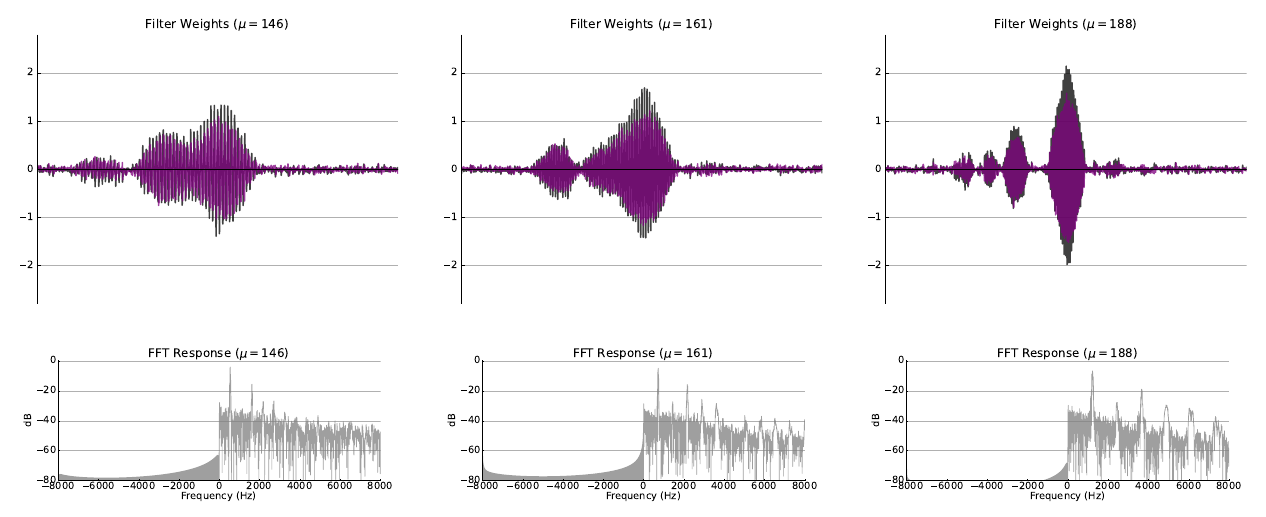}}
\caption{Examples of filters from the \textit{hb+comb+var} experiment with two or more main lobes.}
\end{figure*}

\clearpage

\end{document}